\documentclass[11pt]{article}

\usepackage[margin=1in]{geometry}
\usepackage[T1]{fontenc}
\usepackage{lmodern}
\usepackage{microtype}
\usepackage{xcolor}
\usepackage{color,xspace}
\usepackage[ruled,linesnumbered]{algorithm2e}
\usepackage{float}
\usepackage{graphicx}
\usepackage{cite}
\usepackage{mathtools}
\usepackage{wrapfig}
\usepackage{array}
\usepackage{nicematrix}
\usepackage{tikz}
\usepackage{amsmath}
\usepackage{amssymb}
\usepackage{amsthm}
\usepackage{verbatim}
\usepackage{enumerate}
\usepackage[hidelinks]{hyperref}

\usepackage{thmtools}
\usepackage{thm-restate}
\usepackage[capitalise,nameinlink]{cleveref}

\declaretheorem[name=Theorem]{theorem}
\declaretheorem[name=Lemma,sibling=theorem]{lemma}
\declaretheorem[name=Corollary,sibling=theorem]{corollary}
\declaretheorem[name=Observation,sibling=theorem]{observation}
\declaretheorem[name=Fact,sibling=theorem]{fact}

\declaretheorem[name=Problem,sibling=theorem]{problem}
\declaretheorem[name=Definition,style=definition,sibling=theorem]{definition}

\newcommand{\eg}{e.g.~}

\newcommand{\cube}{\left\lceil(\frac{\sqrt{d}}{2})^d\right\rceil}
\newcommand{\frechet}{Fréchet}

\newcommand{\FD}{\text{FD}_{\delta}}
\newcommand{\curvePoint}{curvePoint}
\newcommand{\side}{\mathrm{side}}
\newcommand{\valid}{valid}

\title{Approximate Map Matching under Realistic Assumptions}

\author{
Lindsey Deryckere\thanks{The University of Sydney, Australia,
\texttt{lindsey.deryckere@sydney.edu.au}}
\and
Joachim Gudmundsson\thanks{The University of Sydney, Australia,
\texttt{joachim.gudmundsson@sydney.edu.au}}
\and
André van Renssen\thanks{The University of Sydney, Australia,
\texttt{andre.vanrenssen@sydney.edu.au}}
\and
Sampson Wong\thanks{The University of Copenhagen, Denmark,
\texttt{sawo@di.ku.dk}}
}

\date{}

\begin{document}
\maketitle

\begin{abstract}
    Given a graph $G=(V,E)$ and a polygonal curve $P$ of complexity $n$ in $\mathbb{R}^d$, we study the map matching problem of computing a path in $G$ that minimises the Fréchet distance to $P$. Chen, Driemel, Guibas, Nguyen and Wenk (ALENEX 2011) gave a near-linear time $(1+\varepsilon)$-approximation for the case when $G$ is $\lambda$-low-dense and $P$ is $c$-packed. We show that these assumptions can be relaxed, and we give efficient $(1+\varepsilon)$-approximation algorithms under the weaker conditions. First, if $P$ is $c$-packed and $G$ is an arbitrary graph, we obtain an $\tilde{O}(cn + c^2|E|)$ time algorithm. Second, if both $G$ and $P$ are $\lambda$-low-dense, we obtain an $\tilde{O}(\lambda^2((n|V|)^{1-1/d}+n+|V|))$ expected time algorithm, which becomes $\tilde{O}(\lambda^2(n+|V|))$ in the plane. 
\end{abstract}


\section{Introduction}
\begin{figure}
    \centering
    \includegraphics[scale=.65]{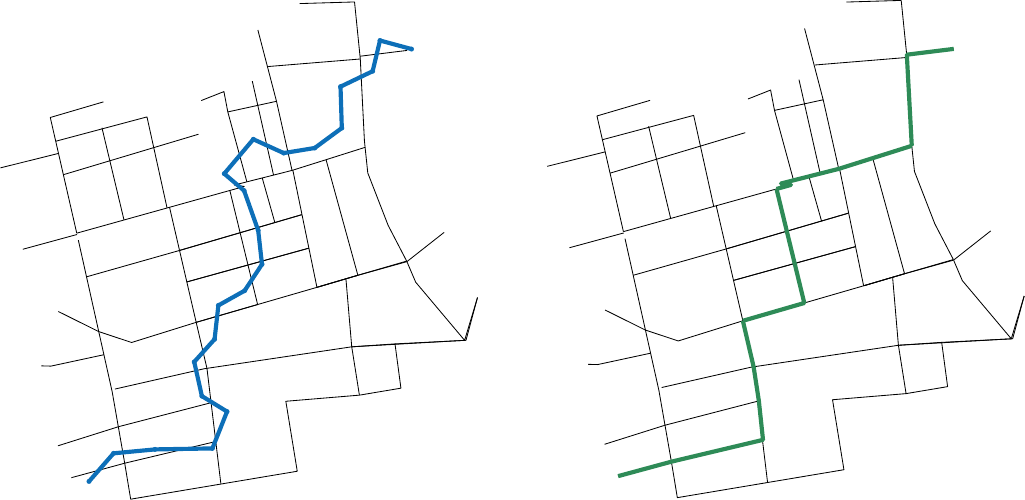}
    \caption{A noisy trajectory (blue), a road network (black) and a map-matched path (green). }
    \label{fig:map_matching_introduce}
\end{figure}

The widespread availability of GPS-enabled devices allows for the continuous collection of vehicle trajectory data, which is widely used in navigation systems, traffic management, and urban planning. However, raw GPS data frequently suffers from measurement noise and low sampling rates, resulting in trajectories that deviate from actual routes. Map matching serves as a fundamental preprocessing task to address this issue by aligning the recorded sequence of locations to a path on a geometric graph representing the road network. The map matching problem has been studied extensively across multiple communities for more than two decades; see the surveys~\cite{DBLP:conf/adc/ChaoXH020,DBLP:journals/urban/HashemiK14,DBLP:journals/itsm/KubickaCMN18,quddus2007current,DBLP:conf/gis/WeiWFZ13,DBLP:books/sp/zheng11}.
Formally, the input to the map matching problem is (i) a trajectory, which represents a vehicle's position over time, and (ii) a network, which represents the roads that the vehicle may have travelled. The output to the map matching problem is a path along the road network that is similar to the input trajectory. See Figure~\ref{fig:map_matching_introduce}. Intuitively, the map-matched path is a ``best guess'' of the path that the vehicle actually travelled.

In theoretical computer science, the most popular approach for the map matching problem is (i) to model the trajectory as a polygonal curve in the Euclidean plane, (ii) to model the network as a graph in the Euclidean plane, and (iii) to measure the similarity of the map-matched path with the input trajectory using some variant of the \frechet{} distance~\cite{NeedMapMatchSpeed, MapMatchVehicleTraj, AltERW03, ChenSW21, ChambersFWW20, GudmundssonS15, GudmundssonSW24, ChenDGNW11, BuchinBG0W24}. Using the standard \frechet{} distance variant offers the advantage of guaranteeing that the matched path is spatially similar to the input trajectory. \newline

\begin{problem}[Map Matching]
    \label{problem:1}
    Given a connected graph~$G$ and a trajectory~$P$, compute a path $\pi \in G$ that minimises $d_F(\pi,P)$, where $d_F$ denotes the Fr\'echet distance. 
\end{problem}

Observe that, under this definition, a path may visit any edge or vertex multiple times but must traverse edges fully. In other words, one cannot partially traverse edges, and changes in direction are only allowed at vertices in the graph. We denote the optimal value $\min_{\pi} d_F(\pi,P)$ of Problem~\ref{problem:1} by $r^*$.

In this paper we mainly consider the decision version of the problem as defined below.
\begin{problem}[Map Matching - Decision Version]
    \label{prob:DecMapMatch}
    Given parameters $\delta \geq 0$, as well as a graph $G$ and polygonal curve $P$ in $\mathbb{R}^d$, does there exist a path $\pi \in G$ such that the \frechet{} distance between $\pi$ and $P$ is at most $\delta$?
\end{problem}

In 2003, Alt, Efrat, Rote and Wenk~\cite{AltERW03} proposed Problem~\ref{problem:1} and provided an $O(mn \log m)$ time algorithm, where $m = |E|+|V|$ is the complexity of the graph~$G = (V,E)$ and $n$ is the complexity of the curve~$P$. One of their key contributions was extending the free-space diagram~\cite{AltG95} to map matching by introducing the \emph{free space surface}.

Conditioned on the Orthogonal Vectors Hypothesis, there exists no $O((mn)^{1-\delta})$ time algorithm for Problem~\ref{problem:1} for any fixed $\delta > 0$~\cite{Bringmann14,BuchinOS19,GudmundssonSW24}.

Although the complexity of Problem~\ref{problem:1} is settled, there has been continued research interest on map matching for classes of \emph{realistic} input graphs and curves. The intuition is that, by applying assumptions that reflect the real-world properties of vehicle trajectories and road networks, one can circumvent the unrealistic lower bounds in~\cite{Bringmann14,BuchinOS19,GudmundssonSW24}. Such approaches have proven successful in the setting of measuring the similarity between curves. One such realistic input model is the concept of $c$-packedness. A set of edges $P$ is $c$-packed if, for every radius $r$ and every ball $B$ of radius $r$, the total length of the edges in $B \cap P$ is at most $c \cdot r$. Driemel, Har-Peled and Wenk~\cite{DriemelHW12} showed that if two input curves satisfy $c$-packedness one can approximate their \frechet{} distance in near-linear time, even in higher dimensions, using simplification techniques. Another realistic model they consider is $\lambda$-low density. A set of edges~$G$ is $\lambda$-low-dense if, for every radius~$r$ and every ball $B$ of radius~$r$, the number of edges in~$G$ that intersect $B$ and have length $\geq r$ is at most $\lambda$. Note that $c$-packedness is a strictly stronger assumption than $\lambda$-low density; any $c$-packed curve is $2c$-low-dense. Nevertheless, Driemel, Har-Peled and Wenk~\cite{DriemelHW12} show that one can still approximate the \frechet{} distance between two $\lambda$-low dense curves in subquadratic time, even in higher dimensions. Diatzko, Lasseux, and Storandt~\cite{diatzko2026} recently established the complementary relationship $c \in O(\lambda\sqrt{n})$ and show the dependence on $n$ is tight.

\begin{table}[t]
\centering
\def\arraystretch{1.2}
        \begin{tabular}{|c|c|c|c|c|c|}
        \hline
         Apx Ratio & Dim & Graph~$G$ & Curve~$P$ & Running time  & Reference  \\
         \hline
         \hline
        $1$ & 2 & any & any & $ \tilde O(|E| \cdot n)$ & \cite{AltERW03} \\
        \hline
         $1+\varepsilon$ & $d$ & $\lambda$-low-dense & $c$-packed & $ \tilde O(\lambda n + c|E|)$ & \cite{ChenDGNW11} \\
         \hline 
         \hline
         $1+\varepsilon$ & $d$ & any & $c$-packed & $ \tilde O(cn+c^2|E|)$& Theorem~\ref{thm:decisionC} \\
             \hline
         $1+\varepsilon$ & $d$ & $\lambda$-low-dense & $\lambda$-low-dense & $\tilde O(\lambda^2((n|V|)^{1-1/d}+n+|V|))$ & Theorem~\ref{thm:low_density}\\
        \hline
        $1+\varepsilon$ & 2 & $\lambda$-low-dense & $\lambda$-low-dense & $ \tilde O(\lambda^2  (n+|V|))$ & Theorem~\ref{thm:low_density}\\
         \hline
    \end{tabular}
    \vspace{3mm}
    \caption{Previous and our results for Problem~\ref{problem:1}. Here, $|E|$ and $|V|$ are the number of edges and vertices of the graph $G$, and $n$ is the complexity of the curve~$P$. $\tilde O(\cdot)$ hides dependence on $\varepsilon$ and polylogarithmic factors, and assumes $d$ is constant. See also the concurrent work in Section~\ref{sec:concurrent}.}
    \label{table:results}
\end{table}

In 2011, Chen, Driemel, Guibas, Nguyen and Wenk~\cite{ChenDGNW11} combined the techniques of~\cite{DriemelHW12} and~\cite{AltERW03} to approximate a solution to Problem~\ref{problem:1} on realistic input graphs and curves. In particular, they assume that the graph~$G$ is $\lambda$-low-dense and the curve~$P$ is $c$-packed. Unlike in~\cite{DriemelHW12} where both input curves are simplified, the authors of~\cite{ChenDGNW11} showed that if one of the curves is $\lambda$-low dense, only the $c$-packed curve requires simplification in order to approximate the \frechet{} distance in subquadratic time. 
They provided a nearly-linear time approximation algorithm for Problem~\ref{problem:1}. Specifically, their $(1+\varepsilon)$-approximation algorithm runs in $O((m+n)\log(m+n) + (\lambda n + cm) \log (mn) \log (m+n) + (\lambda n/\varepsilon^{d} + cm/\varepsilon) \log mn)$ time. 

Gudmundsson, Mai and Wong~\cite{GudmundssonMW24} were the first to consider approximating the \frechet{} distance between two polygonal curves when only one curve is $c$-packed. They obtained a nearly-linear time approximation algorithm that runs in $O(c^3(n + m) \log^{2d+1}(n) \log m)$ time. Conradi, van der Hoog, van der Horst and Ophelders~\cite{DBLP:conf/sosa/ConradiHHO26} improved the runtime to $O(c\frac{n+m}{\varepsilon} \log \frac{n+m}{\varepsilon})$, where $n$ and $m$ denote the complexities of both curves respectively.

In this paper, we consider the following question.

\begin{problem}
    \label{problem:2}
    In~\cite{ChenDGNW11}, are both  assumptions, $G$ is $\lambda$-low-dense and $P$ is $c$-packed, necessary? Can milder assumptions be used to obtain a subquadratic time $(1+\varepsilon)$-approximation algorithm?
\end{problem}

If milder assumptions can be used, then the resulting map matching algorithm can be applied to a wider class of trajectories and road networks. 

\subsection{Contributions}

We prove that, to obtain a subquadratic time $(1+\varepsilon)$-approximation algorithm for Problem~\ref{problem:1}, it is not necessary to assume both $G$ is $\lambda$-low-dense and $P$ is $c$-packed, which answers Problem~\ref{problem:2} in the affirmative.
We provide $(1+\varepsilon)$-approximate deciders for Problem~\ref{prob:DecMapMatch} when:

\begin{enumerate}[(1)]
    \item the graph~$G$ is any general graph, and the curve~$P$ is $c$-packed. In particular, the running time of our decider is $O(cn+\frac{1}{\varepsilon^{2d}}\log^2 \frac{1}{\varepsilon} \cdot n \log^2 n +|V| \frac{1}{\varepsilon^d} (c+\log n)  + |E|\frac{c}{\varepsilon^3}\log\frac{c}{\varepsilon} \cdot \log n \log \log n + |E|\frac{c^2}{\varepsilon^2})$.
    \item the graph~$G$ is $\lambda$-low-dense and the curve~$P$ is $\lambda$-low-dense. In particular after expected $O(\lambda|V|\log |V|)$ preprocessing time, our decider runs in $O\Big(\lambda^2\Big(\frac{(n|V|)^{1-1/d}}{\varepsilon^2}+\frac{n+|V|}{\varepsilon^d}\Big)\log (n|E|)\Big)$ time. The running time is $O\Big(\frac{\lambda^2}{\varepsilon^2}(n+|V|)\log (n|E|)\Big)$ when $d=2$.
\end{enumerate}

If $\varepsilon$ and $d$ are constants, and we ignore polylogarithmic factors in $n$ and~$E$, the running times become $\tilde{O}(cn + c^2|E|)$ and $\tilde{O}(\lambda^2((n|V|)^{1-1/d}+n+|V|))$, respectively. See Table~\ref{table:results}. Using these deciders, we obtain $(1+\varepsilon)$-approximation algorithms for Problem~\ref{problem:1} with the same running times up to logarithmic factors (Section~\ref{app:decider-to-approximation}).

\subsection{Concurrent Work}\label{sec:concurrent}
In a recent preprint, Conradi, van der Hoog, Uldahl, and Rotenberg~\cite{conradi2026tightfrechetboundslambdalow} independently studied Problem~\ref{problem:1} when both $G$ and $P$ are $\lambda$-low-dense. For a curve with $n$ vertices and a graph with $n$ edges, they give a $(1+\varepsilon)$-approximation algorithm that runs in $\tilde{O}(\lambda^{2/d}n^{2-2/d})$ time, and show that for $d\geq 3$ and under the Orthogonal Vectors Hypothesis, no $\tilde{O}((\lambda^{2/d}n^{2-2/d})^{1-\delta})$ time algorithm exists for any $\delta>0$.

\subsection{Other Related Work}

Map matching algorithms under the \frechet{} distance have been studied extensively. However, the question of whether one can preprocess the graph into a data structure that allows for sublinear-time map-matching queries is a more recent line of research. Under the Orthogonal Vectors Hypothesis, no data structure can be constructed in polynomial time to process queries on geometric planar graphs in strongly subquadratic time with respect to the complexity of the input structures~\cite{GudmundssonSW24}. Consequently, research interest has shifted towards applying realistic structural assumptions on the input graph to perform approximate queries.  

Gudmundsson, Seybold and Wong~\cite{GudmundssonSW24} introduced the first query data structure for map matching under the \frechet{} distance under the assumption that the graph is $c$-packed. Under this assumption, their structure uses $\tilde{O}(cm)$ space, and answers $(1+\varepsilon)$-approximate decision queries in $O(c^4 n \log^4 m)$ time, where $n$ is the complexity of the curve, and $m = |V|+|E|$ is the complexity of the $c$-packed graph.

Buchin, Buchin, Gudmundsson, Popov and Wong~\cite{BuchinBG0W24} subsequently extended map matching queries to $\lambda$-low density $t$-spanners, a class of graphs designed to better capture the properties of real-world road networks. Their data structure uses $\tilde{O}(\lambda m^{3/2})$ space and answers approximate \frechet{} distance queries in $\tilde{O}((\lambda+t^2)^4\cdot\lambda^2\cdot n\sqrt{m})$ time. Their data structure also reports the path that $(1+\varepsilon)$-approximates the \frechet{} distance.

\section{Technical overview}
We give a brief overview of our techniques and highlight differences in techniques between this paper and previous results.

\subsection{Map Matching with $c$-packedness}
We first give an overview of our decision procedure and then describe how to apply the framework by~\cite{DriemelHW12} to speed up approximate \frechet{} distance computations by using curve simplification and $c$-packedness. 

\subparagraph{Overview of Approximate Decision Procedure. }
Given the graph $G$ and polygonal curve $P$ in $\mathbb{R}^d$, as well as a query value $\delta \geq 0$, we construct a search structure $H(U,F)$ whose set of nodes $U$ correspond to points (not necessarily vertices) of $P$, and whose set of directed arcs $F$ represent monotone mappings of edges $e$ in $G$ to subtrajectories of $ P$ such that the following two conditions hold:

\begin{enumerate}
    \item If there exists a path in $H$ that starts at a node corresponding to $p_1$ and ends at a node corresponding to $p_n$, then there exists a path $Q \subseteq G$ such that $d_F(Q,P) \leq (1+\varepsilon)\delta$. 
    \item If no such path exists in $H$, then for all paths $Q$ in $G$, $d_F(Q,P) > \delta$.
\end{enumerate}

In other words, we reduce the approximate decision problem to finding a path in $H$. 

We construct $H$ in two stages. In Stage 1 we construct the nodes of $H$. 

Each node corresponds to a sample of the simplified curve $P'$. The sample is a point of a (1D) grid of spacing $\varepsilon' \delta$ along $P'$ that lies on a subtrajectory of $P'$ within distance $r = (1+2\varepsilon')\delta$ of a vertex of $G$. Since a matching may start and end at arbitrary points of $P$, nodes cannot be restricted to vertices of $P$. We preprocess $P'$ in $O(n(\log n + c))$ time into a (push-down) compressed quadtree such that we can efficiently query, for each vertex of $G$, the subtrajectories of $P'$ within distance $r$ of it.

In Stage 2 we construct the directed arcs of $H$. For each edge $e \in G$, we compute which subtrajectories of $P$ can be monotonically mapped to $e$ within \frechet{} distance $(1+\varepsilon)\delta$. For each subtrajectory, if the subtrajectory can be mapped to the edge within \frechet{} distance $(1+\varepsilon)\delta$, we add a directed arc between nodes of $H$ corresponding to samples of the subtrajectory. In this stage we make use of techniques introduced in~\cite{jaywalking} to preprocess the curve $P$ into a data structure that supports efficient queries of the \frechet{} distance between a subtrajectory of the curve and an edge of $G$. Due to the nature of a polygonal curve and the monotonicity of a \frechet{} distance mapping, the resulting structure $H$ is a directed graph. 

Finally, we search $H$ for a path between a node of $H$ corresponding to $p_1$ and a node corresponding to $p_n$. Such a path represents a monotone mapping between a path in $G$ and the entire curve $P$ that satisfies the \frechet{} distance and approximation value requirements.

\subparagraph{Speeding up approximate queries with curve simplification. }
We employ the framework of~\cite{DriemelHW12} to
speed up approximate \frechet{} distance computations by using curve simplification and $c$-packedness. In particular, the authors of~\cite{DriemelHW12} showed that one can simplify a $c$-packed curve $P$ with respect to a parameter $\mu$ in linear time such that the resulting curve $P^\prime$ satisfies:

\begin{itemize}
    \item All except the last segment of $P^\prime$ is guaranteed to have length at least $\mu$.
    \item The \frechet{} distance between $P$ and $P^\prime$ is at most $\mu$.
    \item $P^\prime$ is $6c$-packed. [Lemma 4.3 in~\cite{DriemelHW12}]
\end{itemize}

The above properties of a simplification of a $c$-packed curve allow us to upper bound the number of subtrajectories of $P^\prime$ that can be within distance $\delta$ of a vertex of the graph, hence bounding the complexity of the search structure. Therefore, prior to computing our search structure, we first simplify the $c$-packed curve $P$ with respect to the query value $\delta$ and the approximation value $\varepsilon$. We then construct the search structure $H$ with respect to $P^\prime$ instead of $P$ to reduce the complexity of the search structure and consequently achieve a more efficient runtime. 

\subparagraph{Relation to previous work. }
While our approach builds upon the framework introduced in~\cite{GudmundssonMW24} for approximating the \frechet{} distance between two curves, we adapt and optimise their auxiliary graph construction to accommodate map matching (i.e., matching a c-packed curve against a general graph rather than against a general curve). The authors of~\cite{GudmundssonMW24} compute the approximate \frechet{} distance between a $c$-packed curve $P$ (with $n$ vertices) and a general curve $Q$ (with $m$ vertices). In contrast, our input pairs a $c$-packed curve against a general graph $G=(V,E)$. Our approaches differ as follows.

\emph{Stage 1 (Vertex Construction):} Rather than sampling $O(nc/\varepsilon^2)$ vertices for the auxiliary graph using the range-searching data structure by~\cite{SchwarzkopfV96}, we adapt the data structure by the authors of~\cite{GudmundssonHW23}. This reduces the total vertex count in $H$ to $O(|V|c/\varepsilon)$ and consequently reduces the construction time of the vertices.

\emph{Stage 2 (Edge Construction):} We reduce the number of queries to the data structure by~\cite{jaywalking} and reduce the total edge count from $O(m c^2/\varepsilon^4)$ down to $O(|E|c/\varepsilon)$. Consequently, we reduce the construction time of the edges of $H$.

\emph{Graph Structure and Search:} Because our input is a general graph, the resulting auxiliary graph loses the layered DAG structure present in~\cite{GudmundssonMW24} and becomes a regular directed graph. Consequently, we replace their standard BFS search with a modified DFS search, resulting in an overall running time of $\tilde{O}(cn+c^2|E|)$.

We obtain the following theorem, whose proof is given in Section~\ref{sec:c-packed}.

\begin{restatable}{theorem}{DecisionMapMatchcpacked}
    \label{thm:decisionC}
    Let $P$ be a $c$-packed polygonal curve of complexity $n$, and let $G = (V,E)$ be an arbitrary graph of complexity $|V|+|E|$ in $\mathbb{R}^d$. Given a query value $\delta>0$, and approximation value $0<\varepsilon<1$, 
    there exists an algorithm \textbf{decider} such that, if it returns YES, there exists a path $Q$ in $G$ such that $d_F(P,Q) \leq (1+\varepsilon)\delta$ and, if it returns NO, then for all paths $Q$ in $G$, $d_F(P,Q) > \delta$. The decider runs in $O(cn+\frac{1}{\varepsilon^{2d}}\log^2 \frac{1}{\varepsilon} \cdot n \log^2 n +|V| \cdot \frac{1}{\varepsilon^d} (c+\log n)  + |E|\frac{c}{\varepsilon^3}\log\frac{c}{\varepsilon} \cdot \log n \log \log n + |E|\frac{c^2}{\varepsilon^2})$ time, using $O(cn+|E|\frac{c}{\varepsilon}+(\frac{1}{\varepsilon^{2d}}\log^2 \frac{1}{\varepsilon})n)$ space.
\end{restatable}

\subsection{Map Matching with $\lambda$-Low Density}
Our approach builds upon the framework introduced in~\cite{DriemelHW12} for bounding the approximate free space complexity between two $\lambda$-low density curves, combined with the approximate decider framework for map matching by~\cite{ChenDGNW11} (which builds on~\cite{DriemelHW12} and~\cite{AltERW03}). 

The authors of~\cite{DriemelHW12} introduce a curve simplification algorithm that guarantees two fundamental properties for an input curve $P$:
\begin{enumerate}
    \item All but the last edge has a minimum edge length.
    \item The \frechet{} distance between $P$ and simplified curve $P^\prime$ is bounded relative to this minimum edge length.
\end{enumerate}
While this simplification preserves $c$-packedness up to a factor of 6, it does not similarly preserve the $\lambda$-low density property within a small factor. However, the authors of~\cite{DriemelHW12} observe that if many edges of the simplified $\lambda$-low density curve overlap with a given region (creating high density areas), the original curve must have a lot of vertices nearby to satisfy the low density property of the original curve. They use this observation to charge the number of edges of the simplified curves that contribute to the free space complexity to the original vertices of the graph such that each vertex is only charged a sublinear number of times. Consequently, they show that the free space complexity between the simplified curves is subquadratic in the complexity of the original curves. The charging argument used by the authors of~\cite{DriemelHW12} requires both curves to be simplified as it employs a two-way charging scheme. 

Extending this approach to $\lambda$-low density graphs presents several challenges. While standard graph clustering algorithms can violate either properties of the simplification, they can also significantly inflate local graph density, without guaranteeing that the original graph had many vertices nearby. The main challenge is to produce a simplification whose complexity is small in that of the original graph, while maintaining  the original connectivity of the graph, as well as a small \frechet{} distance to the original graph. 

We propose a two-stage graph simplification scheme. First we use $r$-nets to cluster the graph, enforcing a minimum graph distance between vertices under the original graph metric, while preserving a bounded \frechet{} error. Simplifying the graph however, can significantly increase the local density of the graph. The increase in density can be significantly higher than for curves due to the fact that it needs to maintain more connectivity than a curve. Hence, we apply a second stage of simplification which involves a pruning step. This reduces the density of the graph such that we can bound the number of simplified edges overlapping any region (whose size is related to the query and approximation factor), circumventing the need to link the local density of the simplified graph to the number of vertices of the original graph nearby. Throughout this step, we continue to maintain a bounded \frechet{} distance error. We then use the above techniques, combined with arguments similar to those in~\cite{DriemelHW12}, to bound the free space complexity between the simplified structures, and relate this to the approximate free space between the original structures.

Finally, we apply the framework by~\cite{ChenDGNW11} which uses the approximate free space between the structures to design an efficient approximate decider.

\subsection{From Deciders to Approximation Algorithms}
Following Driemel, Har-Peled and Wenk~\cite[Section~3]{DriemelHW12}, we
turn both deciders into $(1+\varepsilon)$-approximation algorithms for
Problem~\ref{problem:1}, with the same running times up to logarithmic
factors; see Section~\ref{app:decider-to-approximation}.

\section{Preliminaries}
\subsection{Input Model}
We define two properties that relate to realistic graphs and curves. 

\begin{definition}[A $\lambda$-low density collection of edges]
    A collection of edges $G$ in $\mathbb{R}^d$ is considered $\lambda$-low density if, for any ball $B(p,r)$ of any radius $r$, centred at any point $p \in \mathbb{R}^d$, the number of edges of $G$ of length at least $r$ that intersect $B$ is at most $\lambda$.
\end{definition}

\begin{definition}[A $c$-packed curve]
    A polygonal curve $P$ in $\mathbb{R}^d$ is considered $c$-packed if, for any ball $B(p,r)$ of radius $r>0$, centred at any point $p \in \mathbb{R}^d$, the length of the curve that overlaps the ball, i.e., $|B \cap P|$, is at most $c \cdot r$.
\end{definition}

\subsection{The Fr\'echet distance}
\label{sec:problem_definition_free_space}

A polygonal curve $P$ is defined by an ordered set of points $\{ p_1, ..., p_n \} \allowbreak \in \allowbreak \mathbb{R}^d$, called the vertices of $P$. $P$ is a piecewise-linear function $P:[1,n] \rightarrow \mathbb{R}^d$, where for any $t\in[i,i+1]$, $P(t)$ is defined as the linear interpolation of $p_i$ and $p_{i+1}$. In other words, $P(t) = p_i+(t-i)(p_{i+1}-p_i)$. 
A reparameterisation of $[1,n]$ is a non-decreasing surjection $f:[0,1] \rightarrow[1,n]$. Two reparameterisations $f$ and $g$, of $[1,n]$ and $[1,m]$ respectively, describe a matching $(f,g)$ between two curves $P$ and $Q$ with $n$ and $m$ vertices, where for any $t \in [0,1]$, the point $P(f(t))$ is matched to $Q(g(t))$. The continuous \emph{static} \frechet{} distance $d_F(P,Q)$ between two polygonal curves $P$ and $Q$ is defined as  

\begin{align*}
    d_F(P,Q) := \min_{(f,g)}\max_{t\in[0,1]} ||P(f(t)) - Q(g(t))||,
\end{align*}
where $(f,g)$ range over all valid matchings between $P$ and $Q$. Throughout the paper, we denote $d_F(P,Q)$ by $d_F$.\\

Let $G = (V,E)$ be an undirected graph embedded in $\mathbb{R}^d$. Let $V = \{v_1,... v_{|V|}\} \subseteq\mathbb{R}^d$ be the vertices of $G$. Although $G$ is undirected, we regard each undirected edge $(u,v) \in E$ as being represented by two directed edges $(u,v),(v,u) \in E$. Thus, the edges in $E$ are directed but represent an undirected graph. We define a path $\pi$ in the graph to be a polygonal curve formed by its edges. Note that our definition of a path does not restrict $\pi$ from reusing the same edge or vertex multiple times. 

The \frechet{} distance between a curve $P$ and a graph $G$ can now be defined analogously to the case of two curves $P$ and $Q$, if we think of $Q$ as the path in $G$ that minimises the \frechet{} distance to $P$. A common way to visualise the possible reparameterisations of two polygonal curves $P$ and $Q$ is to use the \emph{free space diagram}. The free space diagram ($\FD$) is a geometric object commonly used to decide whether the \frechet{} distance between two polygonal curves $P,Q$ is at most some given parameter $\delta\geq0$. The notion was first introduced in~\cite{AltG95}.

\begin{definition}[Free space]{(From~\cite{AltG95})}
\label{def:freespace}
    Let $P : [0,1] \rightarrow\mathbb{R}^d$, $Q : [0,1] \rightarrow\mathbb{R}^d$ be two curves. The set $F_\delta(P,Q):= \{ (s,t) \in [0,1]^2 | \|P(s) - Q(t) \| \leq \delta\}$ denotes the free space of $P$ and $Q$. We call the partition of $[0,1]^2$ into regions that belong to $F_\delta(P,Q)$ and regions that do not the free space diagram ($\FD(P,Q)$) of $P$ and $Q$. 
\end{definition}

The authors of~\cite{AltG95} showed that $d_F(P,Q) \leq \delta$ if and only if there exists a path from $(0,0)$ to $(1,1)$ in the free space diagram that is monotone in both coordinates and fully contained in $F_\delta(P,Q)$. One can thus decide whether the \frechet{} distance between $P$ and $Q$ is at most $\delta$ by finding a monotone path in $\FD$. 

The authors of~\cite{AltERW03} extended this concept to the setting of map matching. 
Instead of a single flat unit square $[0,1]^2$ to serve as the free space diagram, one constructs an individual free space diagram $[0,1]\times e$ for each edge $e \in E$. Gluing these individual edge diagrams together along the boundaries corresponding to shared vertices in $G$ results in a continuous, branched topological space which they refer to as the \emph{free space surface}.

\subsection{Curve Simplification}
In~\cite{DriemelHW12} the authors introduce a framework for approximately solving the \frechet{} distance problem more efficiently for two curves under realistic input models. They achieve this by simplifying the curves with respect to the approximation factor and bounding the free space between the simplified structures. 

\begin{restatable}[Algorithm 2.1 in~\cite{DriemelHW12}]{fact}{curveSimplification}
    \label{fact:curveSimplification}
    One can simplify any curve $P$ with respect to parameter $\mu$ by scanning through $P$, starting at the first vertex, and at each step only taking the next vertex that is distance at least $\mu$ from the previously taken vertex. The resulting curve $P^\prime$ satisfies:
    \begin{itemize}
        \item All but the last segment of $P^\prime$ has length at least $\mu$.
        \item The \frechet{} distance between the original curve and the simplified curve is at most $\mu$.
    \end{itemize}
    The algorithm runs in linear time.
\end{restatable}

\section{Map Matching with $c$-packedness}
\label{sec:c-packed}
In this section we give a $(1+\varepsilon)$-approximate decider for Problem~\ref{prob:DecMapMatch} in the case when the polygonal curve $P =\langle p_1,...,p_n\rangle$ satisfies $c$-packedness. We place no assumptions on the input graph $G=(V={v_1, …v_{|V|}} ,E)$ and show the properties of the curve are sufficient to design an efficient approximate decider for the problem. 
In this section we prove Theorem~\ref{thm:decisionC}, which we restate below.

\DecisionMapMatchcpacked*

We start with some useful notation.

Let $P = \langle p_1,...,p_n\rangle$ be a polygonal curve in $\mathbb{R}^d$ with total edge length $L = \sum^{n-1}_{i=1}||p_{i+1}-p_i||$. In this section, we view $P$ as a continuous mapping $P:[0,L] \rightarrow \mathbb{R}^d$ parameterised by edge length, where $P(0) = p_1$ and $P(L) = p_n$. For any pair of parameters $s,t \in [0,L]$ with $s \leq t$, we denote by $P[s,t]$ the subtrajectory along $P$ starting at the point $P(s)$ and ending at $P(t)$. Note that the endpoints $P(s)$ and $P(t)$ need not coincide with the vertices of $P$.

\subparagraph{Overview of Approximate Decision Procedure}
Given the graph $G$ and polygonal curve $P$ in $\mathbb{R}^d$, as well as a query value $\delta > 0$ and approximation value $0<\varepsilon<1$, we construct a directed graph $H(U,F)$ whose set of nodes $U$ correspond to points (not necessarily vertices) of $P$, and whose set of directed arcs $F$ represent monotone mappings of edges $e$ in $G$ to subtrajectories $P[s,t] \subseteq P$ such that the following two conditions hold:

\begin{enumerate}
    \item If there exists a path in $H$ that starts at a node corresponding to $p_1$ and ends at a node corresponding to $p_n$, then there exists a path $Q \subseteq G$ such that $d_F(Q,P) \leq (1+\varepsilon)\delta$. 
    \item If no such path exists in $H$, then for all paths $Q$ in $G$, $d_F(Q,P) > \delta$.
\end{enumerate}

We construct $H$ in two stages. In Stage 1 we construct the nodes of $H$. Each node corresponds to a sample point of $P$, on a fixed grid of spacing $\varepsilon' \delta$ along the curve, that lies within distance
$r := (1+2\varepsilon')\delta$ of a vertex of $G$, where $\varepsilon' = \varepsilon /5$.
In Stage 2 we construct the directed arcs of $H$. Each arc in $H$ represents a monotone mapping of a subtrajectory of $P$ to an edge of $G$ such that the \frechet{} distance is at most $\tau=(1+\varepsilon^\prime)r$. 

Finally, we search the resulting graph $H$ for a valid path between a node of $H$ corresponding to $p_1$ and a node corresponding to $p_n$. Such a path represents a monotone mapping between a path in $G$ and the entire curve $P$ that satisfies the \frechet{} distance and approximation value requirements.

In Section~\ref{sec:construction} we describe the construction of our graph $H$ with respect the the graph $G$, $c$-packed curve $P$, our query value $\delta$, and an approximation value $\varepsilon$. In Section~\ref{sec:path_finding} we then show how to use $H$ to answer the decision problem.

\subsection{Constructing the directed graph $H$}
\label{sec:construction}
Prior to computing our graph, we first simplify the $c$-packed curve $P$ with respect to the query value $\delta > 0$ and the approximation value $0<\varepsilon<1$. We then construct the directed graph $H$ with respect to the simplified curve $P^\prime$ instead of $P$ to reduce the complexity of the directed graph and consequently achieve a more efficient runtime. 

We simplify the curve $P$ using the algorithm described in Fact~\ref{fact:curveSimplification} with parameter $\mu = \varepsilon^\prime\delta$, where $\varepsilon^\prime = \varepsilon/5$.  The resulting simplified curve is denoted by $P^\prime$ and satisfies:
\begin{itemize}
    \item All except the last segment of $P^\prime$ is guaranteed to have length at least $\mu$.
    \item The \frechet{} distance between $P$ and $P^\prime$ is at most $\mu$.
    \item $P^\prime$ is $6c$-packed. [Lemma 4.3 in~\cite{DriemelHW12}]
\end{itemize}
 
In the remainder of this section, we use $P^\prime$ to construct our graph $H = (U,F)$. To avoid confusion, we refer to a vertex of $G$ as a \emph{vertex}, while we refer to a vertex of $H$ as a \emph{node}. Similarly, we refer to an edge of $G$ as an \emph{edge}, while we refer to a directed edge of $H$ as a \emph{directed arc}.

\paragraph*{Stage 1: Constructing the Nodes of $H$}
\label{section:push_down_quadtree}

In this section we preprocess $P^\prime$ such that we can efficiently query, for every vertex in $G$, the subtrajectories of $P^\prime$ that are close to that vertex. The grid samples inside such subtrajectories will make up the nodes of $H$. Furthermore, for each vertex $v\in G$ we define its ``$\delta$-neighbourhood'' (or simply ``neighbourhood''), denoted as $\mathcal{N}_v$, as the set of all maximal contiguous subintervals $[s,t] \subseteq [0,L]$ such that $P^\prime[s,t]$ is fully contained within the ball $B(v, r)$, where $r=(1+2\varepsilon')\delta$. The subtrajectories in $\mathcal{N}_v$ are stored in order of occurrence along the curve $P^\prime$ from $p_1$ to $p_n$. To compute the neighbourhoods of vertices in $G$ we first compute the edges of $P^\prime$ that are close to each vertex $v\in G$. We use existing techniques and results by~\cite{GudmundssonHW23} to preprocess $P^\prime$ to enable efficient computation of such edges. Using the edges it is straightforward to determine the neighbourhoods of the vertices in $G$, which we then use to construct the set of nodes $U$ of $H$.

We first describe how to preprocess $P^\prime$ into a quadtree data structure $T$ such that we can efficiently query, for any vertex $v\in G$, which edges of $P^\prime$ lie within distance $r$ of $v$, where $r=(1+2\varepsilon')\delta$.

Recall from Fact~\ref{fact:curveSimplification} that every segment of $P'$, except possibly the last one, has length at least $\varepsilon^\prime\delta$. The last segment is handled separately:
at query time we test it against each query ball directly, in $O(1)$ time per vertex. In the remainder of the construction we therefore assume that every segment of $P'$ has length at least $\varepsilon^\prime\delta$. Let $S$ denote the set of segments of $P'$, so $|S| < n$.

\subparagraph*{Preprocessing.} 
The preprocessing consists of three steps.

\emph{Step 1 (associated canonical hypercubes).} We call a hypercube \emph{canonical} if its side length is a power of $2$ and it is a cell of the hierarchical grid (i.e. a multi-level grid whose cell size is determined by repeatedly halving a larger bounding area) of that side length. Let $\mathcal{B}$ be an axis-aligned bounding box containing $S$, and let $\mathcal{G}$ be a hierarchical grid over $\mathcal{B}$.
For each segment $s \in S$, let $Q(s)$ be the smallest axis-parallel hypercube that is centred at the midpoint of $s$ and contains $s$; its side length, denoted by $\side(Q(s))$, lies in $[\,|s|/\sqrt d,\ |s|\,]$. Using Lemma~9 of~\cite{GudmundssonHW23}, cover $Q(s)$ by $O(1)$ canonical hypercubes of $\mathcal{G}$, each of side length at least $\side(Q(s))/2 \ge |s|/(2\sqrt d)$ and at most $2\,\side(Q(s)) \le 2|s|$.  We say that these canonical hypercubes are \emph{associated} with $s$. Let $\mathcal{C}$ be the set of all associated hypercubes over all segments; $|\mathcal{C}| = O(n)$, and $\mathcal{C}$ is computed in $O(n)$ time.

\emph{Step 2 (compressed quadtree).} Given $\mathcal{B}$, $\mathcal{G}$ and $\mathcal{C}$, build a minimal compressed quadtree $T$ over $\mathcal C$ such that every hypercube of $\mathcal C$ is the region of a node of $T$; the tree $T$ has $O(n)$ nodes and is constructed in $O(n \log n)$ time [~\cite{sariel-book}, Lemma~2.11]. We
further preprocess $T$ in $O(n \log n)$ time using Theorem~2.14 from~\cite{sariel-book} to support \emph{cell queries}: given a canonical hypercube $C$, return the lowest node of $T$ whose region contains $C$, or report that no such node exists, in $O(\log n)$ time.

\emph{Step 3 (push-down).} For a node $u$ of a quadtree, we denote by $R_u$ the hypercube (region) of $u$.
Sort the segments of $S$ in increasing order of length. Each node $u$ of $T$ stores a sorted list $L(u)$ of segments, initially empty. We process the segments in sorted order. To process a segment $s$ in $S$, we locate the nodes of $T$
whose regions are the canonical hypercubes associated with $s$ (these nodes exist by the construction of $T$), and we start a descent at each of them: at a node $u$, if $s \cap R_u \ne \emptyset$, we prepend $s$ to $L(u)$ and recurse into all children of $u$; otherwise we stop. Since segments are processed in increasing order of length, every list $L(u)$ is sorted in decreasing order of segment length, as in~\cite{GudmundssonHW23}.

This completes the construction of $T$.

\subparagraph*{Answering the queries.} Given a query vertex $v \in V$, let $B= B(v, r)$.
Let $b$ be the largest power of $2$ such that $b \le \varepsilon^\prime\delta/(2\sqrt d)$, and consider the canonical grid with cells of side length $b$. Let $\mathcal K_v$ be the set of cells of this grid that intersect $B$; then $|\mathcal K_v| = O\big(((1+2\varepsilon^\prime)\delta/b)^d\big) = O(1/{\varepsilon^\prime}^d)$ for constant $d$, and $\mathcal K_v$ can be enumerated in $O(1/{\varepsilon^\prime}^d)$ time.

For each cell $C \in \mathcal K_v$, we perform a cell query on $T$ to obtain, in $O(\log n)$ time, the lowest node $u_C$ of $T$ whose region contains $C$; if no such node exists we discard $C$ (as argued below, no segment of $S$ intersects such a cell). We then scan the list $L(u_C)$ and, for each of its $O(c)$ segments, test in $O(1)$ time whether it intersects $B$; every segment that does is reported. 

We analyse the construction time and size of $T$.

\begin{lemma}
    \label{lem:quadtreePreprocessSize}
    Given parameters $\delta>0$ and $0<\varepsilon^\prime<\frac{1}{3}$, let $P'$ be a $6c$-packed polygonal curve in $\mathbb{R}^d$ with $n$ vertices, all of whose segments, except possibly the last, have length at least $\varepsilon^\prime\delta$. One can construct $T$ as described above in $O(n (\log n + c))$ time. The size of $T$ is $O(cn)$.
\end{lemma}
\begin{proof}
    We first bound the size of $T$. Observe in step 1 that the set of all associated hypercubes of all segments ($\mathcal{C}$) is of linear size. Therefore, the compressed quadtree is of size at most $O(n)$, as per Lemma~2.11 from~\cite{sariel-book}. It remains to bound, for each $u \in T$ the size of $L(u)$.
    A descent, started at an associated hypercube of $s$, visits only that node and its descendants. Hence a segment $s$ is stored at a node $u$ if and only if $s \cap R_u \ne \emptyset$ and some canonical hypercube associated with $s$ is the region of $u$ or of an ancestor of $u$ in $T$. In particular, every $s \in L(u)$ satisfies
    \begin{equation}
        |s| \ \ge\ \side(R_u)/2,
    \label{eq:long}
    \end{equation}
    since the associated hypercube through which $s$ reached $u$ has side length at least $\side(R_u)$ and at most $2|s|$.
    We can now bound, for all $u \in T$, $|L(u)|$. Let $\ell = \side(R_u)$ and let $z$ be the centre of $R_u$. Every $s \in L(u)$ intersects $R_u$ in some point $q$ and, by~\eqref{eq:long}, has length at least $\ell/2$; hence at least one of the two subsegments of $s$ ending at $q$ has length at least $\ell/4$, and its initial portion of length $\ell/4$ lies within distance $\ell/4$ of $q$, and therefore inside the ball $B(z, \sqrt{d}\,\ell)$. Each segment of $L(u)$ thus contributes length at least $\ell/4$ to $B(z,\sqrt d\,\ell)$. Since $P'$ is $6c$-packed, the total length of $P'$ inside this ball is at most $6c\sqrt d\,\ell$, and consequently,
\begin{equation}
  |L(u)| \ \le\ 24\sqrt d\, c \ =\ O(c).
  \label{eq:listsize}
\end{equation}
    
    It remains to analyse the construction time.
    In Step 1 of the preprocessing $\mathcal{C}$ is computed in $O(n)$ time. In Step 2, $T$ can be constructed in $O(n \log n)$ time, as per Lemma~2.11 in~\cite{sariel-book}. Finally, in Step 3 the segments can be sorted in $O(n \log n)$ time. Locating the $O(1)$ starting nodes of each segment takes $O(\log n)$ time per segment using the structure of Step~2, i.e., $O(n \log n)$ in total. Each insertion takes $O(1)$ time, and by~\eqref{eq:listsize} the total number of insertions is $\sum_u |L(u)| = O(cn)$. Additionally, for every node at which an insertion occurs, the descents inspect its children (of which there are at most $2^d$). Consequently, the total time spent in the descents is $O(cn)$. We conclude the total preprocessing time is $O(n \log n + cn)$.
\end{proof}

Next, we prove the correctness of the data structure and analyse the query time.
\begin{lemma}
    \label{lem:LayeredGraphVertices}
    Given parameters $\delta>0$ and $0<\varepsilon^\prime<\frac{1}{3}$, let $P'$ be a $6c$-packed polygonal curve in $\mathbb{R}^d$ with $n$ vertices, all of whose segments, except possibly the last, have length at least $\varepsilon^\prime\delta$. One can preprocess $P'$ in $O(n(\log n + c))$ time into the data structure of size $O(cn)$ described above such that, given any set $V$ of query points, one can report, for every $v \in V$, all segments of $P'$ that intersect the ball $B(v,r)$, together with, for each segment, the first and last points of each such intersection, in $O\!\big(|V|\frac{1}{{\varepsilon^\prime}^{d}} (c + \log n )\big)$ time.
\end{lemma}

\begin{proof}
We start by proving the correctness of the data structure. 
Every reported segment intersects $B$ by the final test, so it suffices to show that every segment of $S$ intersecting $B$ is tested. Let $s \in S$ with $s \cap B \ne \emptyset$, and let $p \in s \cap B$. Since the cells of $\mathcal K_v$ cover $B$, we have $p \in C$ for some $C \in \mathcal K_v$. Let $Q$ be a canonical hypercube associated with $s$ that contains $p$; such a hypercube exists since the associated hypercubes of $s$ cover $s$. By Step~1 and the minimum segment length of $P'$,
\[
  \side(Q) \ \ge\ \frac{|s|}{2\sqrt d} \ \ge\ \frac{{\varepsilon^\prime}\delta}{2\sqrt d}
  \ \ge\ b \ =\ \side(C).
\]
The canonical hypercubes $Q$ and $C$ both contain $p$, so they are nested, and by the above inequality $C \subseteq Q$. Since $Q$ is the region of a node of $T$ that contains $C$, the cell query for $C$ succeeds and returns a node $u_C$ with $C \subseteq R_{u_C} \subseteq Q$. Every node $w$ on the path in $T$ from $Q$ down to $u_C$ satisfies $p \in C \subseteq R_{u_C} \subseteq R_w$, so $s$ intersects $R_w$; consequently, the descent for $s$ in Step~3, started at $Q$, reached $u_C$ and inserted $s$ into $L(u_C)$. Hence $s$ is tested when the cell $C$ is processed. The same argument shows that if no node of $T$ contains $C$, then no segment of $S$ intersects $C$.

Finally, we analyse the query time. Each vertex of $G$ gives rise to $O(1/{\varepsilon^\prime}^d)$ cells. Each cell requires an $O(\log n)$ time query in $T$. Upon reaching the lowest node in $T$ that contains the desired region, scanning the list of segments stored there takes $O(c)$ time by~\eqref{eq:listsize}. Together with the $O(1)$-time test of the possibly short last segment of $P'$, the time spent per vertex is $O\big(\frac{1}{{\varepsilon^\prime}^d} (c + \log n )\big)$, and the total time over all vertices is $O\big(|V| \cdot \frac{1}{{\varepsilon^\prime}^d} (c + \log n )\big)$.

The size and construction time of $T$ follow from Lemma~\ref{lem:quadtreePreprocessSize}.
\end{proof}

\begin{figure}
    \begin{center}
        \includegraphics[scale=.8]{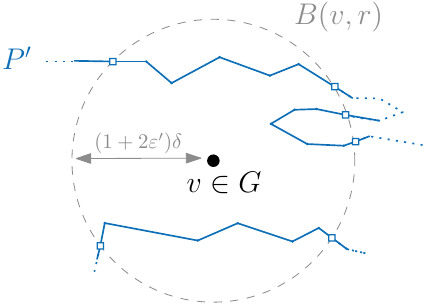}
        \caption{A visualisation of the edges of $P^\prime$ that intersect $B(v,r)$ with $r=(1+2\varepsilon^\prime)\delta$, as well as the intersection between the edges and boundary of the ball. }
        \label{fig:edgesBall}
    \end{center}
\end{figure}

\begin{figure}
    \begin{center}
        \includegraphics[scale=.8]{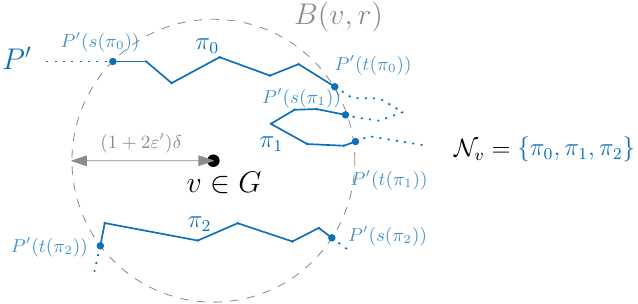}
        \caption{The subtrajectories added to the neighbourhood $\mathcal{N}_v$ and their endpoints on $P^\prime$.}
        \label{fig:neighborhoods}
    \end{center}
\end{figure}

We use the above data structure to query, for every vertex $v \in G$, all edges $e \in P^\prime$ that intersect the ball $B(v,r)$, along with the first and last point along the edge that intersects with the ball (see Figure~\ref{fig:edgesBall}) . 
Let $\mathcal{S}_v$ denote the set of edges that intersect the ball corresponding to $v$. We sort the edges in order of their appearance on $P'$. For each vertex of $G$ we then iterate through the edges in sorted order to find the start and endpoints of the maximal contiguous subtrajectories that lie fully within each ball. We add each subtrajectory to the neighbourhood $\mathcal{N}_v$ of $v$, in order of appearance along $P^\prime$ (see Figure~\ref{fig:neighborhoods}).

We argue, for each $v \in G$, the neighbourhood of $v$ can be constructed in $O(\frac{1}{{\varepsilon^\prime}^d} (c+\log n)+ \frac{c}{\varepsilon^\prime}\log \frac{c}{\varepsilon^\prime})$ time, and each neighbourhood is of size $O(\frac{c}{\varepsilon^\prime})$. 

\begin{lemma}
    \label{lem:size_layer}
    Given parameters $\delta>0$ and $0<\varepsilon^\prime<\frac{1}{3}$, let $P'$ be a $6c$-packed polygonal curve in $\mathbb{R}^d$ with $n$ vertices, all of whose segments, except possibly the last, have length at least $\varepsilon^\prime\delta$. One can preprocess $P'$ in $O(n(\log n + c))$ time into the data structure of size $O(cn)$ described above such that, in $O\left(|V| \cdot ( \frac{1}{{\varepsilon^\prime}^d} (c+\log n)+ \frac{c}{\varepsilon^\prime}\log \frac{c}{\varepsilon^\prime})\right)$ time, one can construct for all $v\in G$, the neighbourhood $\mathcal{N}_v$. The size of $\mathcal{N}_v$ for any $v\in G$ is $O(\frac{c}{\varepsilon^\prime})$.
\end{lemma}
\begin{proof}
    From Lemma~\ref{lem:LayeredGraphVertices} we deduce it takes $O\left(|V| \cdot \left( \frac{1}{{\varepsilon^\prime}^d} (c+\log n) \right)\right)$ time to compute, for every vertex $v \in G$, the set of segments of $P^\prime$, along with their start and endpoints, that intersect $B(v,r)$. Given the set of segments for each vertex, it then takes an additional $O(\frac{c}{\varepsilon^\prime} \log \frac{c}{\varepsilon^\prime})$ time to sort the segments and $O(\frac{c}{\varepsilon^\prime})$ time to compute the start and endpoints of the maximal continuous subtrajectories that lie inside the ball, which make up the neighbourhood $\mathcal{N}_v$ of $v$.

    It remains to prove the size of the neighbourhoods constructed. Since all but the last edge of $P^\prime$ have length at least $\varepsilon^\prime\delta$, any edge that intersects $B(v,r)$ must contribute at least $\varepsilon^\prime\delta$ to $B(v,r+\varepsilon^\prime\delta)$. Because $P^\prime$ is $6c$-packed, it follows there can be at most $\frac{6c(r+\varepsilon^\prime\delta)}{\varepsilon^\prime\delta} = O(\frac{c}{\varepsilon^\prime})$ edges of $P^\prime$ in $B(v,r+\varepsilon^\prime\delta)$ that overlap with $B(v,r)$. 
\end{proof}

\begin{figure}
    \begin{center}
        \includegraphics[scale=.65]{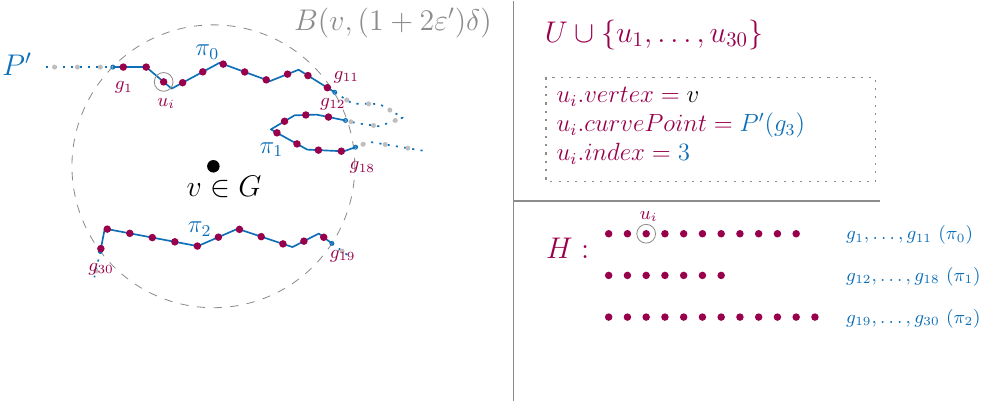}
        \caption{Visualisation of the construction of the vertices of $H$. }
        \label{fig:vertices}
    \end{center}
\end{figure}

We now construct the nodes of $H$ from the neighbourhoods. Let $\Gamma=\{0,\varepsilon'\delta, 2\varepsilon'\delta, \ldots\} \cap [0,L] \cup\{L\}$ be the (1D) grid of spacing $\varepsilon'\delta$ along $P'$. For each vertex $v \in G$, and for each subtrajectory $P'[s,t]$ in $\mathcal{N}_v$, we add one node to $U$ for every grid point $g \in \Gamma\cap [s,t]$, starting at $0$. We call these the samples of $\mathcal{N}_v$, and denote them, in order along $P'$, by $g_1^v < g_2^v < \ldots < g_{m_v}^v$. For each node, we store three attributes. The first attribute, \emph{\curvePoint}, stores the parameter $g$ of the sample, and thus the point $P'(g)$ that the node \emph{corresponds} to\footnote{Note that there can be multiple distinct nodes in $U$ that correspond to the same point along the curve $P^\prime$, \eg if the point lies in both $B(v,r)$ and $B(w,r)$.}. The second attribute, \emph{vertex}, stores the vertex in $G$ that the neighbourhood of the subtrajectory corresponds to. Finally, the last attribute \emph{index}, stores the position $i$ of the sample $g_i^v$ in the ordered list of samples of $\mathcal{N}_v$ (see Figure~\ref{fig:vertices}).

The number of samples of $\mathcal{N}_v$ is $m_v = O(c/\varepsilon')$: for every sample $g$ of $\mathcal{N}_v$ other than $L$, the piece $P'[g, g+\varepsilon'\delta]$ has length $\varepsilon'\delta$ and lies in $B(v, r+\varepsilon'\delta)$, and these pieces are interior-disjoint;
since $P'$ is $6c$-packed there are at most $6c(r + \varepsilon'\delta)/(\varepsilon'\delta) + 1 = O(c/\varepsilon')$ of them. Given the subtrajectories of $\mathcal{N}_v$ in order, the samples are enumerated in $O(|\mathcal{N}_v| + m_v) = O(c/\varepsilon')$ time.

Since the nodes of $H$ are constructed from the samples of the neighbourhoods, it follows that $|U| = O(\sum_{v\in G} m_v)$. Therefore, $H$ has at most $O(|V| \cdot \frac{c}{\varepsilon^\prime})$ nodes. We conclude that after $O(n(\log n +c))$ preprocessing, one can construct both the neighbourhoods of all vertices in $G$, and construct the nodes of $H$ in $O\big(|V| \cdot \left( \frac{1}{{\varepsilon^\prime}^d} (c+\log n) + \frac{c}{\varepsilon^\prime}\log \frac{c}{\varepsilon^\prime}\right)\big)$ time. 
\begin{corollary}
    \label{col:verticesHConstruction}
    Given parameters $\delta>0$ and $0<\varepsilon^\prime<\frac{1}{3}$, let $P'$ be a $6c$-packed polygonal curve in $\mathbb{R}^d$ with $n$ vertices, all of whose segments, except possibly the last, have length at least $\varepsilon^\prime\delta$. One can construct the set of nodes $U$ of $H$ in $O\big(n (\log n+c) + |V|  \left( \frac{1}{{\varepsilon^\prime}^d} (c+\log n) + \frac{c}{\varepsilon^\prime}\log \frac{c}{\varepsilon^\prime}\right)\big)$ time, using $O(cn)$ space. Furthermore, $|U| \leq O(|V|\cdot \frac{c}{\varepsilon^\prime})$.
\end{corollary}

We now use the neighbourhoods of the vertices of $G$ to construct the arcs of $H$. 

\paragraph*{Stage 2: Constructing the Arcs of $H$}
In this section we compute the arcs of $H$ using the neighbourhoods of the vertices in $G$ computed in the previous stage. Recall from above that $\varepsilon^\prime = \varepsilon / 5$, $r=(1+2\varepsilon')\delta$ and $\tau = (1+\varepsilon')r$. Let $u_i,u_j \in U$ be nodes of $H$, and let $s = u_i.\curvePoint \leq u_j.\curvePoint = t$. Let $\pi = P^\prime[s,t]$ be the subtrajectory on $P^\prime$ defined by the endpoints. We add an arc $f = (u_i, u_j)$ to $F$ only if there exists an edge $e \in G$ such that $d_F(\pi, e) \leq \tau$, or a vertex $v \in G$ such that $d_F(\pi, v) \leq r$; exactly which arcs are added is specified below. When this is the case, we say $\pi$ can be \emph{mapped} to edge $e$ (or the degenerate edge $(v,v)$).

Observe that the neighbourhoods of vertices in $G$ already contain all maximal continuous subtrajectories of $P^\prime$ that can be \emph{mapped} to that vertex.
To query whether a subtrajectory $\pi \subseteq P^\prime$ can be \emph{mapped} to an edge $e \in G$, we preprocess $P^\prime$ in $O(\frac{1}{{\varepsilon^\prime}^{2d}}\log^2(\frac{1}{{\varepsilon^\prime}})n \log^2 n)$ time into the data structure by~\cite{jaywalking} (see Fact~\ref{fact:jaywalkingDataStructure}), such that, given any subtrajectory $P^\prime[s,t]$ of $P^\prime$ and any edge $e \in G$, one can compute a value $\delta'$ with $d_F(P^\prime[s,t], e) \leq \delta' \leq (1+\varepsilon') \cdot d_F(P^\prime[s,t], e)$ in $O(\frac{1}{{\varepsilon^\prime}^2} \log n \log(\log n))$ time. 
For each edge in $G$, we use the neighbourhoods corresponding to the endpoints of the edge to determine which subtrajectories of $P^\prime$ to query the above data structure with. To reduce the number of queries, we use the fact that the neighbourhoods are sorted in order of appearance along $P^\prime$, along with the properties of a polygonal curve to employ a binary search. This allows us to construct the set of arcs of $H$ in $O(|E|\frac{c}{{\varepsilon^\prime}^3}\log \frac{c}{{\varepsilon^\prime}} \cdot \log n \log\log n)$ time. We describe our construction in detail below. First, we introduce some notation to help relate points (not necessarily vertices) on $P^\prime$ to nodes of $H$ and vice versa. 

Let $\gamma : ([0,L], V \rightarrow U)$ be a function that maps a sample point to its corresponding node in $H$. For example, $\gamma(g_i^v, v)$ returns the node $u \in U$
\emph{corresponding} to the $i$th sample of the neighbourhood of $v$. In other words, $\gamma(g, v) = u$ if and only if $u.\curvePoint = g$ and $u.vertex = v$. If this is the case, we say node $u\in H$ \emph{corresponds} to point $P'(g)$ on $P^\prime$.

We partition the arcs of $H$ into three types: \emph{traversal}-, \emph{inter-neighbourhood}-, and \emph{intra-neighbourhood} arcs. We describe the construction of each separately.  

\begin{figure}
    \begin{center}
        \includegraphics[scale=.65]{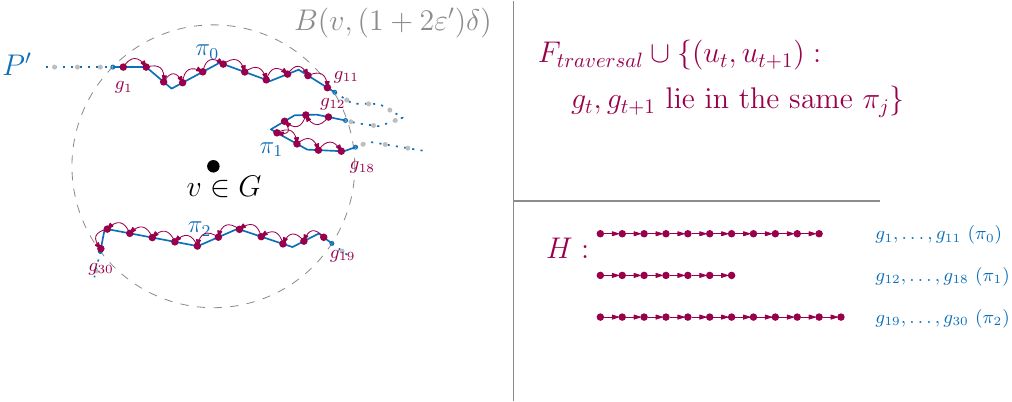}
        \caption{A visualisation of the traversal arcs with respect to the neighbourhood of $v$. Consecutive samples within the same subtrajectory of $\mathcal N_v$ are joined by a traversal arc.}
        \label{fig:traversal_arcs}
    \end{center}
\end{figure}

\subparagraph*{Traversal Arcs}
The purpose of traversal arcs is to represent the potential monotone mapping of a subtrajectory of $P^\prime$ to a vertex $v$ of $G$. Naturally, for each vertex $v \in G$, all subtrajectories $\pi$ in the neighbourhood of $v$ satisfy $d_F(\pi,v) \leq r$. To reflect this in $H$ we add, for each vertex $v\in G$, and for each pair of consecutive samples $g_i^v, g_{i+1}^v$ of $\mathcal{N}_v$ that lie in the same subtrajectory of $\mathcal{N}_v$, a directed arc from $\gamma(g_i^v, v)$ to $\gamma(g_{i+1}^v,v)$, see Figure~\ref{fig:traversal_arcs}.

\subparagraph*{Inter-Neighborhood Arcs}
An inter-neighbourhood arc $(u_i, u_j) \in F$ connects nodes $u_i, u_j \in U$ corresponding to samples in the neighbourhoods of adjacent vertices $(v,w)$ in $G$. Such arcs represent the potential monotone mapping between the subtrajectory $P^\prime[u_i.\curvePoint, u_j.\curvePoint]$ of $P^\prime$ and the corresponding edge $(v,w)\in G$. 

\begin{figure}
    \begin{center}
        \includegraphics[scale=.65]{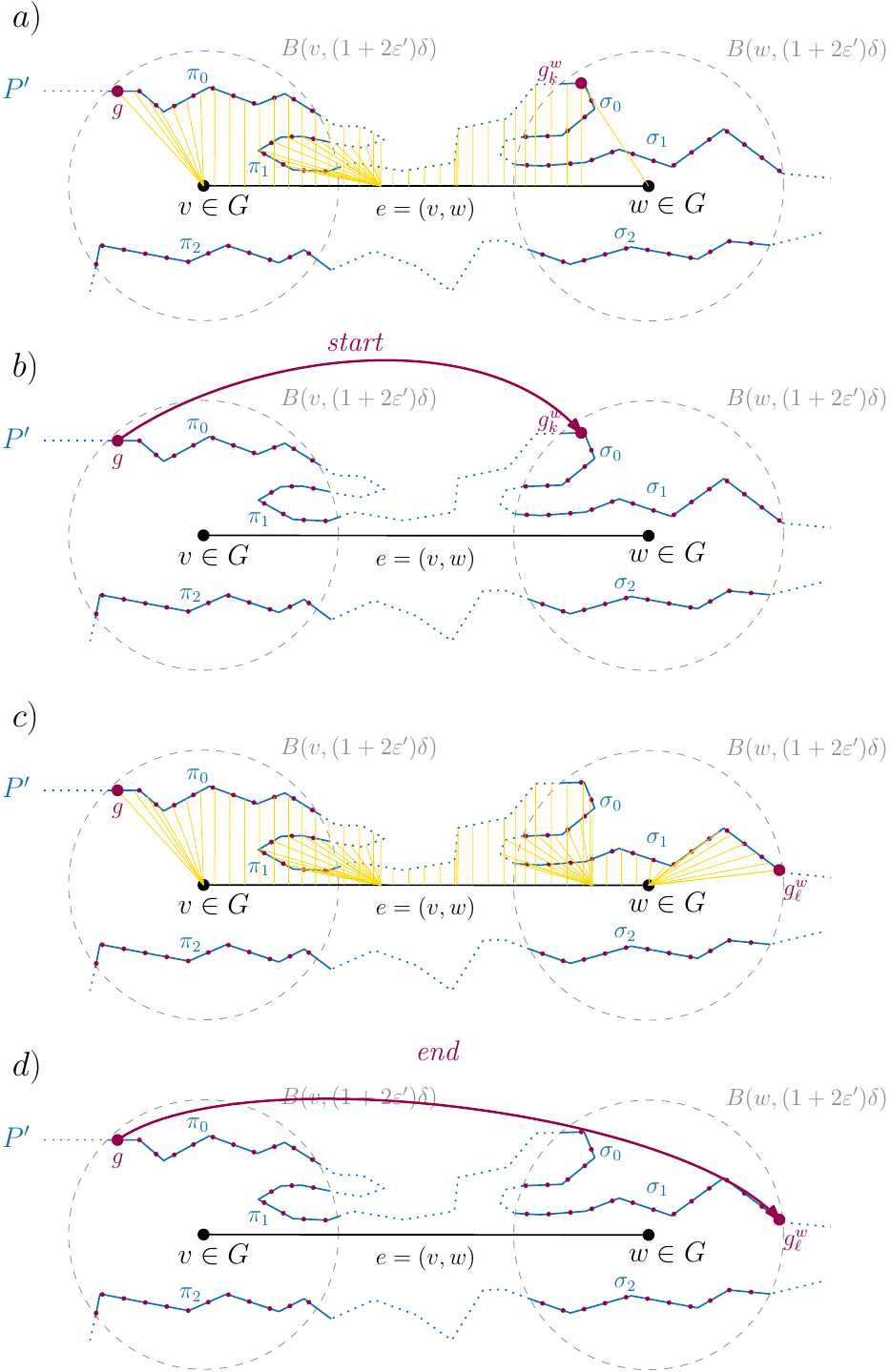}
        \caption{Part a) shows a sample $g$ of $\mathcal N_v$ and the first sample $g^w_k$ of $\mathcal N_w$ with $g^w_k\ge g$ such that $P'[g,g^w_k]$ can be mapped to $e$ within Fréchet distance $\tau$. Part b) shows that consequently we draw an inter-neighbourhood arc, labelled \emph{start}, from $g$ to $g^w_k$. Similarly, part c) shows that $g^w_\ell$ is the last sample of $\mathcal N_w$ such that $P'[g,g^w_\ell]$ can be mapped to $e$ within Fréchet distance $\tau$, and part d) shows the resulting inter-neighbourhood arc, labelled \emph{end}, from $g$ to $g^w_\ell$.}
        \label{fig:inter_neighborhood}
    \end{center}
\end{figure}

For each edge $e=(v,w) \in G$, let $\mathcal{N}_{v}$ and $\mathcal{N}_{w}$ be the neighbourhoods corresponding to its endpoints. For each sample $g$ of $\mathcal{N}_v$, we aim to find the first\footnote{``First'' here is defined with respect to the ordering of the samples of $\mathcal{N}_w$.} sample $g_k^w$ of $\mathcal{N}_w$ with $g_k^w \geq g$ such that $d_F\big(P^\prime[g, g_k^w], (v,w)\big) \leq \tau$. If this is the case, $g$ can reach $g_k^w$. Once found, we add a directed arc from $\gamma(g,v)$ to $\gamma(g_k^w, w)$ in $H$. We label the arc as \emph{start} to indicate that this is the \emph{first} sample of $\mathcal{N}_w$ that $g$ can \emph{reach}. We then repeat the process to find the last sample $g_{\ell}^w$ of $\mathcal{N}_w$, $k\leq \ell$, such that $d_F(P^\prime [g, g_\ell^w], (v,w)) \leq \tau$. We label such arcs with an \emph{end} label (see Figure~\ref{fig:inter_neighborhood}). Note that $g_k^w=g$ is allowed. The edge $(v,w)$ is then traversed while the curve waits at $P'(g)$, which is possible whenever $P'(g)$ is within distance $r$ of both $v$ and $w$. Consequently, $H$ may contain directed cycles of arcs between nodes corresponding to the same parameter. Precisely how $k$ and $\ell$ are computed, and the guarantees they satisfy, are given in Corollary~\ref{cor:binsearch} below.

Although $G$ is undirected, we assume each undirected edge $(v,w)$ is represented by two directional edges; $(v,w)$ and $(w,v)$. Hence, we perform each query twice (once for $(v,w)$ and once for $(w,v)$). 

To determine whether $d_F(P^\prime [g, g_\ell^w], (v,w)) \leq \tau$ efficiently we preprocess $P^\prime$ into the data structure by~\cite{jaywalking}.
\begin{fact}[Theorem 5.9 in~\cite{jaywalking}]
    \label{fact:jaywalkingDataStructure}
    Let $0<\varepsilon^\prime<\frac{1}{3}$. Given a polygonal curve $P^\prime$ with $n$ vertices in $\mathbb{R}^d$, one can build a data structure in $O(\frac{1}{{\varepsilon^\prime}^{2d}}\log^2 \frac{1}{{\varepsilon^\prime}} \cdot n \log^2 n)$ time, that uses $O(\frac{1}{{\varepsilon^\prime}^{2d}}\log^2\frac{1}{{\varepsilon^\prime}} \cdot n)$ space, such that for a query segment $(v,w)$, and any two points $P^\prime(a), P^\prime(b)$ on the curve (and the segments of the curve that contain them), one can $(1+{\varepsilon^\prime})$-approximate the \frechet{} distance $d_F(P^\prime[s,t], (v, w))$ in $O(\frac{1}{{\varepsilon^\prime}^2}\log n \log \log n)$ time.
\end{fact}

For each sample $g_i^w$ of $\mathcal{N}_w$ one can query the data structure with the edge $(v,w)$, and the subtrajectory $P^\prime[g,g_i^w]$ to determine the $(1+\varepsilon^\prime)$-approximate \frechet{} distance. If the approximate \frechet{} distance, $\delta^\prime$, output by the data structure is at most $\tau$, we say $g$ can \emph{reach} $g_i^w$. We further reduce the number of queries using the observation that the neighbourhoods store the samples in sorted order. Instead of, for each sample $g$ of $\mathcal{N}_v$, querying all samples of $\mathcal{N}_w$ to find the first and last sample that $g$ can reach, we employ a binary search on the samples of $\mathcal{N}_w$ (on their order along $P'$) to find, for each sample of $\mathcal{N}_{v}$, the first and last sample of $\mathcal{N}_{w}$ that it can \emph{reach}. The binary search is justified by the following monotonicity property.

\begin{observation}\label{obs:truncate}
Let $e=(a,b)$, $\rho>0$ and $s\le x\le y$ with $d_F(P'[s,y],e)\le\rho$.
(i) If $\|P'(x)-b\|\le\rho$ then $d_F(P'[s,x],e)\le\rho$.
(ii) If $\|P'(x)-a\|\le\rho$ then $d_F(P'[x,y],e)\le\rho$.
\end{observation}
\begin{proof}
In a matching in which every matched pair is at distance at most $\rho$, $P'(x)$ is matched to some point $e(\mu_x)$. Since
$\|P'(x)-e(\mu)\|$ is convex in $\mu$, it is at most $\rho$ on the whole subsegment between
$e(\mu_x)$ and $b$ (for (i)), respectively $a$ (for (ii)). Keep the matching on one side of
$(P'(x),e(\mu_x))$ and match $P'(x)$ to the rest of $e$.
\end{proof}

Fix a sample $g$ of $\mathcal{N}_v$ and the directed edge $e=(v,w)$, and write
$g_j=g^w_j$ for the samples of $\mathcal{N}_w$, $1\le j\le m_w$. For $g_j\ge g$, write
$D(j)=d_F(P'[g,g_j],e)$, and let $\delta'_j$ be the value returned by the data structure
of Fact~13 for $P'[g,g_j]$ and $e$, so that $D(j)\le\delta'_j\le(1+\varepsilon')D(j)$.
Let $k$ be the smallest index with $g_k\ge g$; it is found by a binary search on the samples
of $\mathcal{N}_w$. We then find $\ell$ by a binary search over $\{k-1,\dots,m_w+1\}$,
maintaining indices $\mathit{lo}<\mathit{hi}$ with the invariants that $\mathit{lo}=k-1$ or
$\delta'_{\mathit{lo}}\le\tau$, and that $\mathit{hi}=m_w+1$ or $\delta'_{\mathit{hi}}>\tau$.
Initially $\mathit{lo}=k-1$ and $\mathit{hi}=m_w+1$; in each step we query the middle index
and update $\mathit{lo}$ or $\mathit{hi}$ accordingly, until $\mathit{hi}=\mathit{lo}+1$. We
set $\ell=\mathit{lo}$. If $\ell\ge k$, we add the \emph{start} arc $(\gamma(g,v),\gamma(g_k,w))$
and the \emph{end} arc $(\gamma(g,v),\gamma(g_\ell,w))$; otherwise we add no arc.

\begin{corollary}\label{cor:binsearch}
With the notation above, the search uses $O(\log\frac{c}{\varepsilon'})$ queries, and
\begin{enumerate}
  \item[(i)] $D(j)\le\tau$ for all $k\le j\le\ell$, and
  \item[(ii)] $D(j)>r$ for all $j>\ell$ (note that $j>\ell\ge k-1$ implies $j\ge k$, so $D(j)$
  is defined).
\end{enumerate}
\end{corollary}

\begin{proof}
The number of queries follows from $m_w=O(c/\varepsilon')$.

For (i), let $k\le j\le\ell$. By the invariant, $D(\ell)\le\delta'_\ell\le\tau$. Moreover,
$g\le g_j\le g_\ell$ and $\|P'(g_j)-w\|\le r\le\tau$, since $g_j$ is a sample of
$\mathcal{N}_w$. Observation~\ref{obs:truncate}(i) with $\rho=\tau$ gives $D(j)\le\tau$.

For (ii), let $h=\ell+1\le m_w$. By the invariant, $\delta'_h>\tau$, so
$D(h)\ge\delta'_h/(1+\varepsilon')>\tau/(1+\varepsilon')=r$. Suppose that $D(j)\le r$ for
some $j>h$. Since $g\le g_h\le g_j$ and $\|P'(g_h)-w\|\le r$, Observation~\ref{obs:truncate}(i)
with $\rho=r$ gives $D(h)\le r$, a contradiction.
\end{proof}

\begin{figure}
    \begin{center}
        \includegraphics[scale=.65]{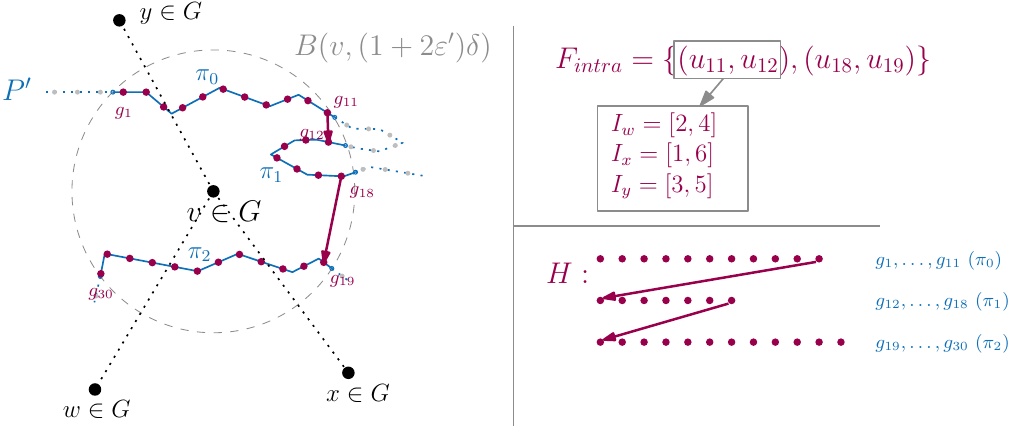}
        \caption{A visualisation of the intra-neighbourhood arcs, which join the last sample of a subtrajectory of $\mathcal N_v$ to the first sample of the next one, and examples of their ranges with respect to each neighbour of $v$.}
        \label{fig:intra_arcs}
    \end{center}
\end{figure}

\subparagraph*{Intra-Neighborhood Arcs}
Intra-neighbourhood arcs connect samples within the same neighbourhood. These arcs are necessary to limit the inter-neighbourhood arcs constructed above to two per sample (a \emph{start} and an \emph{end} arc). Intra-neighbourhood arcs connect the last sample of a subtrajectory of a neighbourhood to the first sample of the subsequent subtrajectory of that neighbourhood that contains a sample. In other words, for all $v \in G$, for every pair of consecutive samples $g_i^v, g_{i+1}^v$ of $\mathcal{N}_v$ that lie in different subtrajectories of $\mathcal{N}_v$, we construct a directed intra-neighbourhood arc from $\gamma(g_i^v,v)$ to $\gamma(g_{i+1}^v, v)$ (see Figure~\ref{fig:intra_arcs}). 

Each intra-neighbourhood arc $(\gamma(g_i^v,v), \gamma(g_{i+1}^v, v))$ in the neighbourhood of a vertex $v \in G$ stores, for each adjacent vertex $w \in G$, an interval $I_w = [a,b]$ of indices of samples of $\mathcal{N}_w$ such that $d_F(P'[g_x^w, g_{i+1}^v], (w,v)) \leq \tau$ for every $x \in I_w$; how $I_w$ is chosen is described below.

To compute the ranges for each intra-neighbourhood arc we use the information from the incoming inter-neighbourhood arcs while traversing the samples of every neighbourhood in order. For each neighbourhood $\mathcal{N}_{v}$ with samples $g_1^v, \ldots , g_{m_v}^v$, the samples are processed in order of appearance\footnote{Recall that these are already stored in sorted order of appearance along $P^\prime$.} along $P^\prime$. During the traversal we keep track, for each neighbouring vertex $w \in G$, of the set of \emph{active} samples of $\mathcal{N}_{w}$. We say a sample $g_x^w$ of $\mathcal{N}_{w}$ is \emph{active} if we have encountered a sample $g_i^v$ of $\mathcal{N}_{v}$ during our traversal, which corresponds to the endpoint of an inter-neighbourhood arc $(\gamma(g_x^w, w), \gamma(g_i^v, v)) \in H$ with a start label, originating from $g_x^w$, and have not yet encountered an inter-neighbourhood arc with the corresponding end label.
We process each sample $g_i^v$ of $\mathcal{N}_v$ as follows. For each neighbouring vertex $w \in G$, we first check the set of inter-neighbourhood arcs coming into $\gamma(g_i^v, v)$. For any arc with a start label we update the relevant set of active samples for the neighbourhood corresponding to the start point of the edge. We then update the range $I_w$ of the intra-neighbourhood arc entering $\gamma(g_i^v,v)$, if there is one, for each neighbouring vertex $w\in G$, to include the indices of the current set of active samples of $\mathcal{N}_w$. Finally, if any of the inter-neighbourhood arcs coming into $\gamma(g_i^v,v)$ have an end label, originating from $g_x^w$, we \emph{de-activate} the sample by removing its index from the relevant range of active samples.  
We store as $I_w$ the smallest interval containing the indices of the active samples of $\mathcal{N}_w$ (maintained with a priority queue with lazy deletions). Although the active set itself need not be contiguous when the queries are approximate, every index in $I_w$ is valid: let $x \in I_w$, so that there are active samples $g_y^w, g_z^w$ with $y\leq x \leq z$. As $g_y^w$ is active, Corollary~\ref{cor:binsearch} (applied to $g_y^w$ and the edge $(w, v)$) gives $d_F(P'[g^w_y, g^v_i]$, $(w, v)) \leq \tau$, $g^w_x \leq g^w_z \leq g^v_i$, and $||P'(g^w_x) - w|| \leq r \leq \tau$. Observation~\ref{obs:truncate}(ii) therefore gives $d_F(P'[g^w_x,g^v_i],(w, v)) \leq \tau$.

This completes the construction of $H$. Next, we argue the correctness of the structure and analyse its size and construction time. For ease of notation, we say a path in $H$ is \emph{\valid{}} if it starts at a node $u_i\in H$ such that $u_i.\curvePoint = 0$ and ends at a node $u_j$ such that $u_j.\curvePoint = L$, and every intra-neighbourhood arc on it satisfies the condition of Section~\ref{sec:path_finding} with respect to the last inter-neighbourhood arc preceding it on the path. 

\begin{lemma}
    \label{lem:constructHtime}
    Given parameters $\delta>0$ and $0<\varepsilon^\prime<\frac{1}{3}$, let $P$ be a $c$-packed curve of complexity $n$ and $G(V,E)$ be a general graph in $\mathbb{R}^d$. One can construct a directed graph $H$ of size $O(|E|\frac{c}{{\varepsilon^\prime}})$ in $O(\frac{1}{{\varepsilon^\prime}^{2d}}\log^2 \frac{1}{{\varepsilon^\prime}} \cdot n \log^2 n +cn+|V|\frac{1}{\varepsilon'^d}(c+\log n) + |E|\frac{c}{{\varepsilon^\prime}^3}\log \frac{c}{{\varepsilon^\prime}} \cdot \log n \log\log n)$ time, using $O(cn+ |E|\frac{c}{{\varepsilon^\prime}}+(\frac{1}{{\varepsilon^\prime}^{2d}}\log^2 \frac{1}{{\varepsilon^\prime}})n)$ space, such that, if there exists a \valid{} path in $H$, then there exists a path $Q \in G$ such that $d_F(Q,P)\leq (1+\varepsilon)\delta$, and if no valid path exists, then for all paths $Q\in G$, $d_F(Q,P)> \delta$. 
\end{lemma}
\begin{proof}
For a node $u$ we call $u.\curvePoint$ its \emph{parameter}. Traversal and intra-neighbourhood arcs join a sample of some $\mathcal{N}_v$ to the next sample of $\mathcal{N}_v$, and an inter-neighbourhood arc from $\gamma(g,v)$ ends at a node of parameter at least $g$. Hence the parameters along any path in $H$ are non-decreasing. Recall that $\tau=(1+\varepsilon')r$ with $r=(1+2\varepsilon')\delta$ and $\varepsilon'=\varepsilon/5$, and that $d_F(P,P')\le\varepsilon'\delta$.

\emph{Soundness.} Let $W$ be a \valid{} path, and let $(\gamma(a_i,v_{i-1}),\gamma(b_i,v_i))$, $1\le i\le m$, be its inter-neighbourhood arcs in order. Then $Q=(v_0,\dots,v_m)$ is a path in $G$. Set $a_{m+1}=L$. Between two consecutive inter-neighbourhood arcs, $W$ stays in $\mathcal{N}_{v_i}$, moving from $b_i$ to $a_{i+1}$ along traversal and intra-neighbourhood arcs. Let $c_i$ be the target
of the last intra-neighbourhood arc on this part, or $c_i=b_i$ if there is none. Then $P'[c_i,a_{i+1}]$ is covered by traversal arcs, so it lies in a single subtrajectory of $\mathcal{N}_{v_i}$ and hence in $B(v_i,r)$. Likewise, since $W$ is \valid{}, $P'[0,a_1]$ lies in $B(v_0,r)$. Moreover, $d_F(P'[a_i,c_i],(v_{i-1},v_i))\le\tau$. If $c_i=b_i$, this holds by Corollary~\ref{cor:binsearch}(i), since $b_i$ is the $k$th or $\ell$th sample found for $a_i$. Otherwise, the intra-neighbourhood arc entering $\gamma(c_i,v_i)$ has a range containing the index of $a_i$, and the bound holds by the validity of the ranges shown above. Matching each piece $P'[c_i,a_{i+1}]$ to $v_i$ and each piece $P'[a_i,c_i]$ to the edge $(v_{i-1},v_i)$, consecutive matchings agree at their common endpoint. This gives $d_F(Q,P')\le\tau$, and therefore $d_F(Q,P) \le \tau+\varepsilon'\delta = (1+4\varepsilon' + 2\varepsilon'^2)\delta \le (1+\varepsilon)\delta$.

\emph{Completeness.} Let $Q=(v_0,\dots,v_m)$ be a path in $G$ with $d_F(Q,P)\le\delta$, so $d_F(Q,P')\le(1+\varepsilon')\delta$. Fix such a matching, and choose parameters $0=y_0\le y_1\le\dots\le y_m=L$ with $P'(y_i)$ matched to $v_i$. For $1\le i\le m$ we then have $d_F(P'[y_{i-1},y_i],(v_{i-1},v_i))\le(1+\varepsilon')\delta$. Let $g_i$ be the largest grid point of $\Gamma$ with $g_i\le y_i$; thus $g_0=0$, $g_m=L$ and $g_0\le\dots\le g_m$. Since $P'[g_i,y_i]$ has length at most $\varepsilon'\delta$, it lies in $B(v_i,r)$. Hence $g_i$ is a sample of $\mathcal{N}_{v_i}$, and prepending $P'[g_{i-1},y_{i-1}]$, matched to $v_{i-1}$, gives $d_F(P'[g_{i-1},y_i],(v_{i-1},v_i))\le r$. Observation~\ref{obs:truncate}(i) with $\rho=r$ and $x=g_i$ then yields $d_F\big(P'[g_{i-1},g_i],(v_{i-1},v_i)\big)\le r$.

Let $k$ and $\ell$ be the indices computed for the sample $g_{i-1}$ and the edge $(v_{i-1},v_i)$, and let $j_i$ be the index of $g_i$ in $\mathcal{N}_{v_i}$. Then $j_i\ge k$ because $g_i\ge g_{i-1}$, and $j_i\le\ell$ by Corollary~\ref{cor:binsearch}(ii). So $H$ contains the \emph{start} arc from $\gamma(g_{i-1},v_{i-1})$ to $\gamma(g^{v_i}_k,v_i)$. Consecutive samples of $\mathcal{N}_{v_i}$ are joined by a traversal or an intra-neighbourhood arc. The index of $g_{i-1}$ is active from $g^{v_i}_k$ to $g^{v_i}_\ell$, so it lies in the range of every intra-neighbourhood arc entering a sample $g^{v_i}_{j'}$ with $k<j'\le j_i$. Hence $\gamma(g_i,v_i)$ is reached from $\gamma(g_{i-1},v_{i-1})$ along a \valid{} path. Concatenating these paths for $i=1,\dots,m$ yields a \valid{} path from $\gamma(0,v_0)$ to $\gamma(L,v_m)$. If $m=0$, then $P'$ lies in one subtrajectory of $\mathcal{N}_{v_0}$, and the traversal arcs lead from $\gamma(0,v_0)$ to $\gamma(L,v_0)$.

\emph{Construction time and space.} By Corollary~\ref{col:verticesHConstruction}, the neighbourhoods and the $O(|V|\frac{c}{\varepsilon'})$ nodes of $H$ are computed in $O\big(n(\log n+c)+|V|\big(\frac{1}{\varepsilon'^d}(c+\log n) +\frac{c}{\varepsilon'}\log\frac{c}{\varepsilon'}\big)\big)$ time and $O(cn)$ space. Since $m_v=O(\frac{c}{\varepsilon'})$, there are $O(|V|\frac{c}{\varepsilon'})$ traversal and intra-neighbourhood arcs, and they are built in the same time. The data
structure of Fact~\ref{fact:jaywalkingDataStructure} takes $O(\frac{1}{\varepsilon'^{2d}}\log^2\frac{1}{\varepsilon'}\cdot n\log^2 n)$ time and $O(\frac{1}{\varepsilon'^{2d}}\log^2\frac{1}{\varepsilon'}\cdot n)$ space. For each of the $\sum_v m_v\deg(v)=O(|E|\frac{c}{\varepsilon'})$ pairs of a sample and an incident directed edge, Corollary~\ref{cor:binsearch} uses $O(\log\frac{c}{\varepsilon'})$ queries and produces at most two inter-neighbourhood arcs. This takes $O(|E|\frac{c}{\varepsilon'^3}\log\frac{c}{\varepsilon'}\cdot\log n\log\log n)$ time in total and gives $O(|E|\frac{c}{\varepsilon'})$ arcs. Finally, each of the $O(\frac{c}{\varepsilon'}\deg(v))$ inter-neighbourhood arcs entering $\mathcal{N}_v$ causes $O(1)$ priority-queue operations, and each of the $O(\frac{c}{\varepsilon'})$ intra-neighbourhood arcs of $\mathcal{N}_v$ stores $\deg(v)$ ranges. The ranges therefore take $O(|E|\frac{c}{\varepsilon'}\log\frac{c}{\varepsilon'})$ time and $O(|E|\frac{c}{\varepsilon'})$ space. Summing these bounds gives the claimed time and space.
\end{proof}

\subsection{Finding a Path in $H$}
\label{sec:path_finding}
It remains to show how to determine whether there exists a \valid{} path in $H$. Whether an
intra-neighbourhood arc may be used depends on the last inter-neighbourhood arc taken before
it, so we search over pairs consisting of a node and this context. Recall that $H$ may
contain directed cycles, all of them formed by inter-neighbourhood arcs between nodes with
the same parameter.

For a node $u$ and a node $a$ (or $a=\bot$), we call the pair $(u,a)$ a \emph{state}. It
represents being at $u$ when the last inter-neighbourhood arc taken started at $a$, where
$a=\bot$ means that no inter-neighbourhood arc has been taken yet. From a state $(u,a)$ the
following transitions are allowed:
\begin{itemize}
  \item along a traversal arc $(u,u')$, to the state $(u',a)$;
  \item along an intra-neighbourhood arc $(u,u')$, to the state $(u',a)$, provided
        $a\ne\bot$ and $a.index$ lies in the range $I_{a.vertex}$ stored at the arc;
  \item along an inter-neighbourhood arc $(u,u')$, to the state $(u',u)$.
\end{itemize}
By definition, a path in $H$ is \valid{} if and only if it corresponds to a sequence of
allowed transitions that starts at a state $(u,\bot)$ with $u.\curvePoint=0$ and ends at a
state $(u',a)$ with $u'.\curvePoint=L$.

To test the range condition in $O(1)$ time, every inter-neighbourhood arc for an edge
$(w,v)$ stores the position of this edge in the adjacency list of $v$, and every
intra-neighbourhood arc of $\mathcal{N}_v$ stores its ranges in an array indexed by these
positions. A state therefore also carries this position, which we omit from the notation.

\subparagraph*{The search.}
We maintain a stack $\mathcal{S}$ of states, and for every node $u$ a flag
$\mathit{expanded}(u)$, initially false.
\begin{enumerate}
  \item \emph{Initialisation.} For every node $u$ with $u.\curvePoint=0$, push $(u,\bot)$
        onto $\mathcal{S}$.
  \item While $\mathcal{S}$ is not empty, pop a state $(u,a)$.
  \begin{enumerate}
    \item If $u.\curvePoint=L$, return YES.
    \item If $u$ has an outgoing traversal arc $(u,u')$, push $(u',a)$.
    \item If $u$ has an outgoing intra-neighbourhood arc $(u,u')$, $a\ne\bot$, and
          $a.index\in I_{a.vertex}$, push $(u',a)$.
    \item If $\mathit{expanded}(u)$ is false, set it to true and push $(u',u)$ for every
          outgoing inter-neighbourhood arc $(u,u')$.
  \end{enumerate}
  \item Return NO.
\end{enumerate}
The outgoing inter-neighbourhood arcs of a node are expanded only once. This suffices
because the state reached through an inter-neighbourhood arc $(u,u')$ is always $(u',u)$,
no matter which state at $u$ we came from. Storing with every pushed state the state from
which it was pushed allows us to report a \valid{} path, and hence the path $Q$ in $G$,
whenever the search returns YES.

\begin{lemma}
    \label{lem:findPathH}
    The search returns YES if and only if $H$ contains a \valid{} path, and it runs in
    $O\big(|V|\frac{c}{\varepsilon'}+|E|\frac{c^2}{\varepsilon'^2}\big)$ time.
\end{lemma}

\begin{proof}
Every pushed state is reached from an initial state by allowed transitions, so the search
returns YES only if a \valid{} path exists.

Conversely, call a state \emph{reachable} if it can be reached from an initial state by
allowed transitions. We show that every reachable state is pushed. Initial states are
pushed in the initialisation step. Suppose a reachable state $(u,a)$ is pushed; then it is
also popped, unless the search has already returned YES. When $(u,a)$ is popped, its
successors along traversal and intra-neighbourhood arcs are pushed in steps (b) and (c).
Its successors along inter-neighbourhood arcs are the states $(u',u)$ for the arcs
$(u,u')$, which do not depend on $a$. These are pushed when $u$ is expanded, which happens
at the latest when $(u,a)$ is popped. Hence, if a \valid{} path exists, a state $(u',a')$
with $u'.\curvePoint=L$ is pushed, and the search returns YES when it is popped.

For the running time, observe that every node of $\mathcal{N}_v$ has at most one outgoing
traversal or intra-neighbourhood arc, and this arc leads to the next sample of
$\mathcal{N}_v$. Hence the states pushed in steps (b) and (c) form chains: each chain starts
at an initial state or at a state pushed in step (d), and visits the samples of a single
neighbourhood in increasing order. A chain therefore has length at most
$m_v=O(\frac{c}{\varepsilon'})$. There are at most $|V|$ initial states, one per vertex
whose ball contains $p_1$. Each inter-neighbourhood arc is followed in step (d) at most
once, because each node is expanded at most once, so there are at most
$O(|E|\frac{c}{\varepsilon'})$ states pushed in step (d). Every pop takes $O(1)$ time apart
from the expansion in step (d), and the expansions take $O(|E|\frac{c}{\varepsilon'})$ time
in total. Summing up, the search takes
$O\big((|V|+|E|\frac{c}{\varepsilon'})\cdot\frac{c}{\varepsilon'}\big)
=O\big(|V|\frac{c}{\varepsilon'}+|E|\frac{c^2}{\varepsilon'^2}\big)$ time. In particular,
the search terminates even though $H$ may contain cycles.
\end{proof}

Combining Lemmas~\ref{lem:constructHtime} and~\ref{lem:findPathH} allows us to conclude
Theorem~\ref{thm:decisionC}.

\section{Map Matching with Low-Density}
\label{sec:lowdense}
In this section we give a $(1+\varepsilon)$-approximate decider for Problem~\ref{prob:DecMapMatch} in the case when both the input graph and the polygonal curve satisfy $\lambda$-low density. Our decider runs in $O\Big(\lambda^2\Big(\frac{(n|V|)^{1-1/d}}{\varepsilon^2}+\frac{n+|V|}{\varepsilon^d}\Big)\log (n|E|)\Big)$ expected time. Note that, although we used $\pi$ to denote a subtrajectory of a polygonal trajectory $P$ above, in this section we use $\pi$ to denote a path in a graph.

\begin{restatable}{theorem}{lowdense}
 \label{thm:low_density}
    Given parameters $\delta>0$ and $0<\varepsilon<1$, as well as a $\lambda$-low density graph $G(V,E)$ of complexity $|V|+|E|$, and a $\lambda$-low density curve $P$ of complexity $n$ in $\mathbb{R}^d$. One can decide in expected $O\Big(\lambda^2\Big(\frac{(n|V|)^{1-1/d}}{\varepsilon^2}+\frac{n+|V|}{\varepsilon^d}\Big)\log (n|E|)\Big)$ time,  whether there is a path $\pi$ in $G$ such that $d_F(P,\pi) \leq (1+\varepsilon)\delta$ or whether, for all paths $\pi$ in $G$, $d_F(P,\pi) > \delta$. In the case where $d=2$ the bound simplifies to $O\Big(\frac{\lambda^2}{\varepsilon^2}(n+|V|)\log (n|E|)\Big)$.
\end{restatable}

We achieve the results by simplifying both structures with respect to the parameters $\delta$ and $\varepsilon$, and showing that the free-space complexity between the simplified structures is subquadratic. Here, following Alt, Efrat, Rote and Wenk~\cite{AltERW03}, the free space is the set of pairs of points, one on the curve and one on the graph, at distance at most $\delta$. The free space surface consists of one cell for each pair of an edge of the curve and an edge of the graph, and the free-space complexity is the number of cells that contain free space. We then relate the existing results from~\cite{ChenDGNW11} and \cite{DriemelHW12} to show that this implies the existence of an approximate decider for the map matching problem that runs in subquadratic time. In particular, given values $\delta$ and $\varepsilon$, we show one can preprocess $G$ and $P$ into simplified structures $G^\prime$ and $P^\prime$ in expected $O(\lambda |V|\log |V|)$ time, such that the free space between the two structures is subquadratic. We then use this free space to design an approximate decider for the original structures that decides whether there exists a path $\pi \in G$ such that the \frechet{} distance between $\pi$ and $P$ is at most $(1+\varepsilon)\delta$ or whether, for all paths $\pi$ in $G$, $d_F(P,\pi) > \delta$. If such a path exists then the decider outputs it.

The authors of~\cite{DriemelHW12} show a linear time algorithm to simplify a polygonal curve $P$ with respect to parameter $\mu = \varepsilon'\delta$ (Fact~\ref{fact:curveSimplification}). 
However, simplifying a low-density graph such that the desired properties of the graph are preserved requires more work. We show in Section~\ref{sec:simplifyGraph} how to simplify the graph to obtain useful properties for bounding its free space complexity. 
First, we revisit some existing results that will be useful for our analysis.

Throughout this section we will use $C$ to denote a hypercube, and $xC$ to denote the scaling of $C$ by a factor of $x$ around its centre. 
\subsection{Useful facts}
We use the curve simplification of Driemel, Har-Peled and Wenk~\cite{DriemelHW12} (Fact~\ref{fact:curveSimplification}).

We will use the following fact to bound the number of edges of length $\ell$ of a $\lambda$-low density graph that can intersect a hypercube of side length $\ell$. To simplify the arguments, we will work with hypercubes rather than balls throughout this section. 

\begin{fact}[Claim 4.7 in~\cite{DriemelHW12}]
\label{fact:cube}
    Let $G$ be a $\lambda$-low-density graph, and let $C$ be a hypercube in $\mathbb{R}^d$ with side length $\ell$. The number of edges of length at least $\ell$ of $G$ that intersect $C$ is bounded by $\lambda \cube$.
\end{fact}

As shown in Lemma~4.9 in~\cite{DriemelHW12}, in order for a $\lambda$-low dense structure to have a lot of edge length within an area, there must exist a lot of vertices in a nearby area. We will require this result to relate the free space complexity between the simplified graph and the simplified curve to the number of vertices in the original structure. However, the original proof of Lemma~4.9 in~\cite{DriemelHW12} assumes $\lambda$ to be a constant and does not state the dependency on the parameter. In this paper we do not consider $\lambda$ to be constant. For this reason we include the full proof with dependency on $\lambda$ below. We use that every vertex of a $\lambda$-low density graph has degree at most $\lambda$.

\begin{restatable}[cf. Lemma 4.9 in~\cite{DriemelHW12} with dependence on $\lambda$]{lemma}{fortynine}
    \label{lem:49}
    Let $G$ be a $\lambda$-low dense graph or path in $\mathbb{R}^d$, and let $C$ be a hypercube with side length $\ell$. Let $\alpha$ be the length of the intersection between $G$ and $C$, and let $M$ be the number of vertices of $G$ contained in $3C$. Then $M=\Omega\Big(\Big(\frac{\alpha}{\lambda\ell}\Big)^{1+\frac{1}{d-1}}\Big)$,  which is equivalent to $\alpha = O(\lambda \ell M^{1-{1/d}})$. 
\end{restatable}

\begin{proof}
We first observe that every vertex $u$ of $G$ has degree $O(\lambda)$. Fix a $\rho$, $0<\rho\le\ell$, and partition the edges of $G$ that intersect $C$ into two sets depending on their length. An edge of length less than $\rho$ that intersects $C$ has both endpoints within distance $\rho\le\ell$ of $C$, hence, they must lie in $3C$. Since every vertex has degree $O(\lambda)$, there are $O(\lambda M)$ such edges, and together they contribute $O(\lambda M\rho)$ to $\alpha$.

Next, cover $C$ by $O((\ell/\rho)^d)$ cubes of side length $\rho$. By Fact~\ref{fact:cube}, each of these cubes meets at most $\lambda\lceil(\frac{\sqrt d}{2})^d\rceil=O(\lambda)$ edges of length at least $\rho$, and each such edge contributes at most $\sqrt d\cdot \rho$ to the small cube. Hence, the edges of length at least $\rho$ contribute
$O(\lambda(\ell/\rho)^d\rho)=O(\lambda\ell^d/\rho^{d-1})$ to $\alpha$, so in total $\alpha = O(\lambda(M\rho+\frac{\ell^d}{\rho^{d-1}}))$.

Choosing $\rho=\ell/M^{1/d}\le\ell$ gives
$\alpha=O\big(\lambda\ell M^{1-1/d}\big)$. Rearranging the terms gives the first statement.
\end{proof}

\subsection{Preprocessing the low-density graph}
\label{sec:simplifyGraph}
The purpose of simplifying the $\lambda$-low density graph $G$ is to reduce the number of edges of the simplified graph that are within distance $\delta$ of edges of $P$, which consequently reduces the free space complexity between the two structures. The aim is to achieve this property while maintaining that for any path in the original graph, there is a path in the simplified graph that is ``close'' with respect to our \frechet{} distance query value. The authors of~\cite{DriemelHW12} showed it is possible to simplify curves such that the simplified curve is close to the original curve in linear time (see Fact~\ref{fact:curveSimplification}). Furthermore, they showed that for a $\lambda$-low density curve in particular, the complexity of the simplification of Fact~\ref{fact:curveSimplification} in any area must be low compared to the complexity of the original curve in a nearby area. This is because the simplification can only violate low density in areas where the curve originally had a lot of edge length, and must therefore have had a lot of vertices in order to maintain low density. One can therefore distribute the charge of the complexity of the simplified structure in this area among the vertices of the original curve in a nearby area such that no vertex is charged more than a sublinear number of times.

Simplifying a low density graph on the other hand is not as straightforward. One approach to reducing the complexity of the graph is to cluster it. However, standard clustering approaches do not achieve the same relation between the complexity of the new graph and the number of vertices of the original graph nearby. This is because the clustered graph must maintain the same connectivity as the original graph in order to maintain that for any path in the original graph, there is a path in the simplified graph that is ``close''. Consequently, the number of edges of the clustered graph in a particular area may significantly increase, without guarantee that the original graph has many vertices in the area, among which the charge can be distributed.

In this section we show how to construct a simplification $G^\prime=(V^\prime, E^\prime)$ of the $\lambda$-low density graph $G$ with respect to the query value of the decision problem $\delta$, and a parameter $\ell$, which is related to the approximation factor $\varepsilon$ and $\delta$, such that the number of edges of the simplified graph that are within distance $\delta$ of an edge of $P$ is at most sublinear in the complexity of $G$. In particular, we process $G$ into $G^\prime$ such that the following properties hold: \vspace{0.2cm}
\begin{description}
    \item[Property 1:] For all paths $\pi \in G$, there exists a path $\sigma \in G^\prime$ such that the \frechet{} distance between $\pi$ and $\sigma$ is at most $(1+4\sqrt{d})\ell$.  Conversely, for all paths $\sigma \in G'$, there exists a path $\pi\in G$ such that the Fréchet distance between $\pi$ and $\sigma$ is at most $\ell$.
    \item[Property 2:] The graph distance in $G$ between vertices of $G^\prime$ connected by an edge in $E^\prime$ is at least $\ell$. In other words, $\forall v^\prime \neq u^\prime \in V^\prime, d_G(u^\prime, v^\prime) \geq \ell$. Note that this does not imply that $G^\prime$ has a minimum edge length.
    \item[Property 3:] Let $C$ be a cube of side length $\delta$. $G^\prime$ cannot have more than $O\Big(\lambda \Big( \frac{|V\cap 27C|^{1-1/d}}{\varepsilon}+ \frac{|V\cap 27C|^{1/d}}{\varepsilon^{d-1}}\Big)\Big)$ edges that overlap with $3C$.
\end{description} \vspace{0.2cm}
We construct $G^\prime$ with respect to the two parameters $\delta$ and $\ell$ in two stages. Throughout both stages we maintain Property~1. Property~2 will be achieved by clustering the graph into an intermediate graph $G^*$. However, clustering the graph may significantly increase the number of simplified edges that overlap with an area, thus violating Property~3. To achieve the third property we therefore prune the clustered graph.

\paragraph*{Stage 1: Clustering the $\lambda$-low density graph}
The purpose of this stage is to reduce the number of small edges in the graph and establish a minimum graph distance (where graph distance is measured with respect to the original graph $G$) between the vertices. Note however that this may not result in a minimum edge length in the clustered graph $G^*$.

To cluster the graph we make use of $r$-nets. An $r$-net of $G$ is a subset $\mathcal{C} \subseteq V$ that satisfies the following two properties:\vspace{0.2cm}
    \begin{description}
        \item [\emph{Separation}] No two points of $\mathcal{C}$ are within graph distance $r$ of one another. 
        \item [\emph{Covering}] Every point in $V$ is within graph distance $r$ of a point of $\mathcal{C}$. 
    \end{description}
    
One can adapt Lemma 2.4 in~\cite{rnets} to $\lambda$-low density graphs by observing that $|E| \leq \lambda |V|$ to obtain:
\begin{fact}[Lemma 2.4 in~\cite{rnets}]
    \label{fact:rnet}
    Given a $\lambda$-low density graph $G= (V,E)$ one can compute an $r$-net of $G$ in $O(\lambda|V| \log |V|)$ expected time.
\end{fact}

We use the randomised construction of~\cite{rnets} to construct an $r$-net with $r$ equal to the specified parameter $\ell$.  Let the resulting $r$-net be $\mathcal{C} = \{c_1,...,c_{|\mathcal{C}|}\}$. We will denote these as cluster centres. We construct corresponding clusters $\mathbb{C} = \{C_1,...,C_{|\mathcal{C}|}\}$ by assigning each vertex of $G$ to its nearest cluster centre, where ``nearest'' is with respect to graph distance in $G$ (computed by a single multi-source run of Dijkstra’s algorithm from all cluster centres in $O(|E| + |V | \log |V |)$ time. By the covering property every vertex is within graph distance $\ell$ of its nearest cluster centre. 

We construct the clustered graph $G^*(V^*, E^*)$ from $\mathcal{C}$ and $G$ as follows. Set $V^* = \mathcal{C}$. We add an edge between $c_i$ and $c_j$ to $E^*$ if there exists an edge in $G$ between any vertex in $C_i$ and any vertex in $C_j$. We store one such edge of $G$ as the witness of $(c_i,c_j)$. Observe that for any path in the $\lambda$-low density graph, there exists a path in the clustered graph such that it is ``close'' with respect to $\ell$.

\begin{observation}
\label{obs:frechetclosepostcluster}
    For any path $\pi \in G$ there exists a path $\sigma \in G^*$ such that $d_F(\pi,\sigma) \leq \ell$. 
\end{observation}
The observation trivially follows from the covering property of $r$-nets, with $r = \ell$. 

\begin{restatable}{observation}{clusterconverse}\label{obs:cluster-converse}
For any path $\sigma\in G^*$ there exists a path $\pi\in G$ such that $d_F(\pi,\sigma)\le\ell$.
\end{restatable}
\begin{proof}
Recall that every vertex of $G$ is assigned to its nearest cluster centre with respect to graph distance in $G$, and that, by the covering property of the $\ell$-net, every vertex of $G$ is within graph distance at most $\ell$ of its cluster centre. Moreover, for every edge $(c_i,c_j)$ of $G^*$ we store a witness edge $(a,b)\in E$ with $a\in C_i$ and $b\in C_j$.

Let $(c_i,c_j)$ be an edge of $\sigma$ with witness $(a,b)$. Let $\pi_a$ be a shortest path in $G$ from $c_i$ to $a$, and let $\pi_b$ be a shortest path in $G$ from $b$ to $c_j$. Both paths have length at most $\ell$, so every point in $\pi_a$ is within Euclidean distance $\ell$ of $c_i$, and every point of $\pi_b$ is within Euclidean distance $\ell$ of $c_j$. We match the walk $\pi_a\circ ab\circ\pi_b$ in $G$ to the segment $c_ic_j$ as follows:
\begin{itemize}
  \item all points of $\pi_a$ are matched to $c_i$;
  \item the edge $ab$ is linearly matched to the segment $c_ic_j$, since $\|a-c_i\|\le\ell$ and $\|b-c_j\|\le\ell$, the convexity of the Euclidean norm implies that every matched pair is at distance at most $\ell$;
  \item all points of $\pi_b$ are matched to $c_j$.
\end{itemize}
This is a monotone matching in which every matched pair is at distance at most $\ell$, so
$d_F(\pi_a\circ ab\circ\pi_b,\, c_ic_j)\le\ell$.

Concatenating these walks over the edges of $\sigma$ yields a walk $\pi$ in $G$. Consecutive
walks meet at a cluster centre, which is matched to itself at the end of one piece and at the
start of the next, so the matchings concatenate and $d_F(\pi,\sigma)\le\ell$.
\end{proof}

After clustering the density of the graph may have significantly increased in certain areas with high connectivity. To limit the number of edges that overlap with an area we prune the resulting graph. 

\paragraph*{Stage 2: Pruning the clustered graph}
The purpose of this stage is to reduce highly connected components of the graph that lie within a small area (small with respect to the parameter $\ell$). As a result, we can bound the number of edges in any area of the pruned graph whose size is related to $\ell$, while maintaining that for any path in the original graph, there is a path that is close to it in the pruned graph.

\begin{figure}
    \centering
    \includegraphics[width=0.4\linewidth]{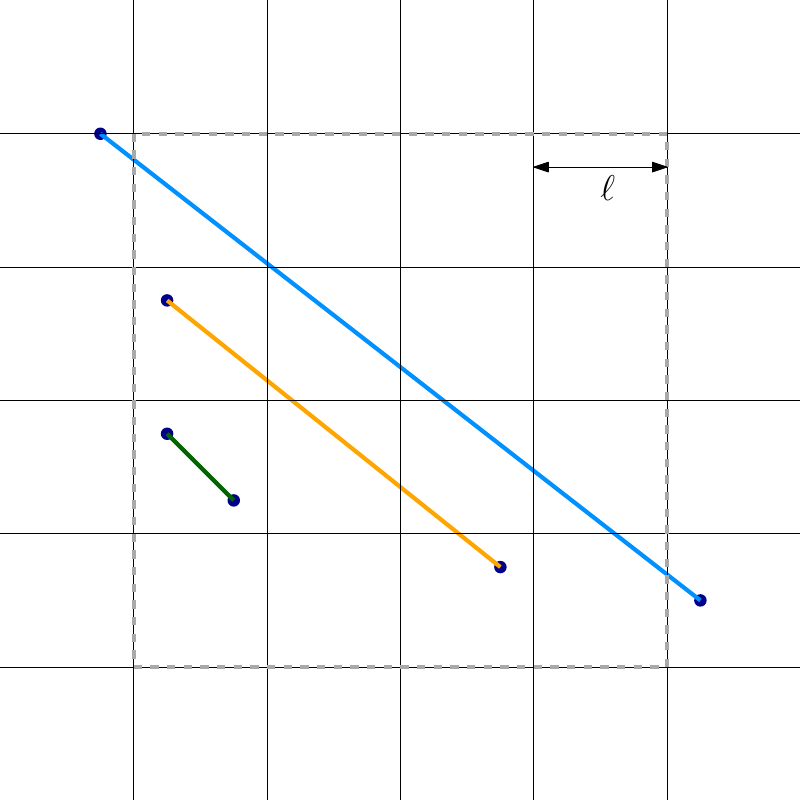}
    \caption{Example of an $\ell$-grid. The green edge is a local edge. The orange edge is a short crossing edge. And the blue edge is a long crossing edge. }
    \label{fig:terminology}
\end{figure}

We start with some terminology. Consider a grid $\mathcal{G}$ of cell size $\ell$ overlaying the $\lambda$-low density graph $G$. Throughout the remainder of Section~\ref{sec:lowdense} we use $\square_\tau$ to denote a block of $\underbrace{\tau \times \tau \times ... \times \tau}_{d}$ cells positioned anywhere in a $d$-dimensional grid of any pre-specified cell size. For $X \subseteq V^*$, let $G^*[X]$ be the subgraph induced by $X$.  
Let $cell(v)$ denote the cell that contains the vertex $v$. We partition the set of edges $E^*$ into three types (see Figure~\ref{fig:terminology} for a visualisation of each type):
\begin{description}
    \item[Local edges:] We call an edge $(u,v)$ local if $cell(u) = cell(v)$. 
    \item[Short crossing edges:] An edge is a short crossing edge if $cell(u) \neq cell(v)$ and the cells belong to a common $\square_4$ block.
    \item[Long crossing edges:] An edge is a long crossing edge if $cell(u) \neq cell(v)$ and the cells do not belong to a common $\square_4$ block.
\end{description}

\begin{figure}
    \centering
    \includegraphics[width=0.7\linewidth]{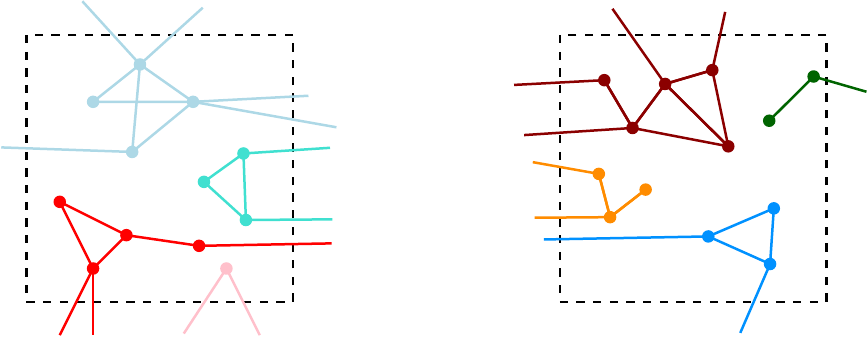}
    \caption{Two cells of the $\ell$-grid, where each cell component of $G$ has been assigned a different colour within each cell.}
    \label{fig:cell_comp}
\end{figure}

\begin{figure}
    \centering
    \includegraphics[width=0.7\linewidth]{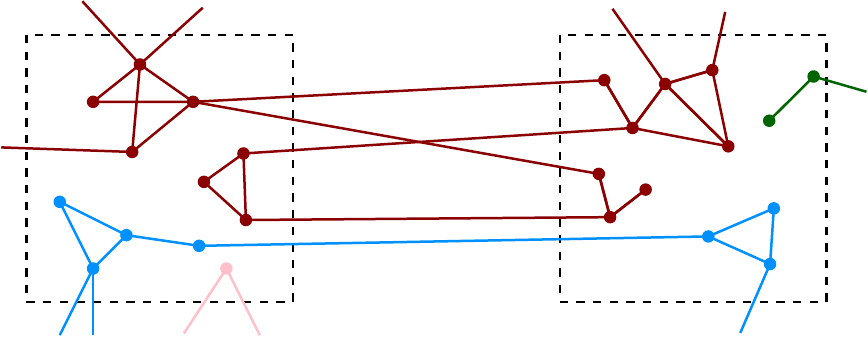}
    \caption{A pair of cells of the $\ell$-grid that lie in a common $\square_4$. Each pair component of $G$ is assigned a different colour. Note that multiple cell components from Figure~\ref{fig:cell_comp} have merged into a single pair component.}
    \label{fig:pair_comp}
\end{figure}

Finally, let $s$ be a cell of the grid. We say a subgraph of $G^*$ is a \emph{cell component} of $s$ if it is a maximal connected component of $G^*[V^* \cap s]$. For a vertex $v \in V^*$ (see Figure~\ref{fig:cell_comp}), let $\tau(v)$ denote its cell component. We say a subgraph of $G^*$ is a \emph{pair component} of a pair of cells ${s,s^\prime} \in \mathcal{G}$ if it is a maximal connected component of $G^*[V^* \cap (s \cup s^\prime)]$ and $s$ and $s^\prime$ lie in a common $\square_4$ block (see Figure~\ref{fig:pair_comp}). Note that every cell component of $s$ or $s^\prime$ is contained in exactly one pair component of $\{s, s^\prime\}$.

We are now ready to describe the pruning procedure to obtain the pruned graph $G^\prime$. The procedure consists of two steps. 

\paragraph*{Step 1 (cells). } For every cell $s$ that contains a vertex, and for every cell component $K$ of $s$, compute a spanning tree $T_K$ of $K$. Remove all edges of $K$ that do not belong to $T_K$. 

\begin{figure}
    \centering
\includegraphics[width=0.75\linewidth]{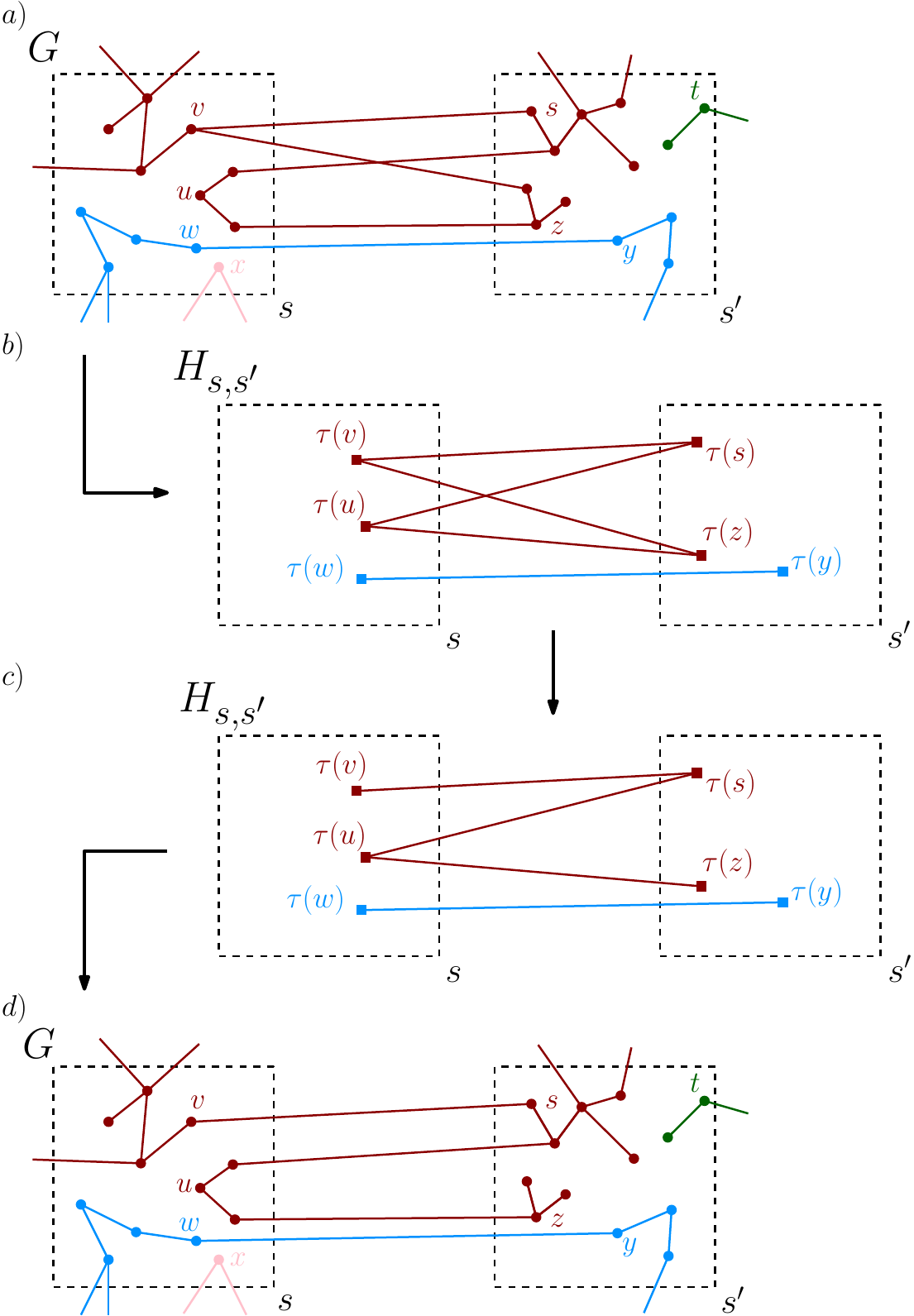}
    \caption{A visualisation of Step 2 of the pruning procedure.}
    \label{fig:prune_pair_components}
\end{figure}

\paragraph*{Step 2 (pairs of cells that share a $\square_4$ block). } For every pair of cells $\{s,s^\prime\}$ that lie in a common $\square_4$ block, let $E^*_{s,s'}$ be the set of short crossing edges with one endpoint in $s$ and the other in $s'$. If $E^*_{s,s'}=\emptyset$, we are done. Otherwise, construct a multigraph $H_{s,s'}$ as follows. The nodes of $H_{s,s'}$ are the cell components of $s$ and $s'$ that contain an endpoint of an edge of $E^*_{s,s'}$. For every edge $(u,v)\in E^*_{s,s'}$, add an edge between $\tau(u)$ and $\tau(v)$ in $H_{s,s'}$. Compute a spanning forest $F_{s,s'}$ of $H_{s,s'}$. Keep exactly the edges of $E^*_{s,s'}$ whose copies lie in $F_{s,s'}$ and delete the other edges of $E^*_{s,s'}$. See Figure~\ref{fig:prune_pair_components}.

We do not remove any long crossing edges of $E^*$. The resulting pruned graph $G^\prime$ consists of the vertex set $V^*$, and its edge set consists of the edges of $E^*$ retained in Steps 1 and 2, together with all long crossing edges in $E^*$.

This completes the construction of the simplified graph $G^\prime$.

We make the following observations, which follow from the construction described above.
From Step 1 it follows that
\begin{observation}
\label{obs:local_forest}
    For every cell $s$, the graph $G'[V^*\cap s]$ is a forest, and its components have the same vertex sets as the cell components of $s$.
\end{observation}
From Step 2 it follows that
\begin{observation}
\label{obs:pair_forest}
    For every pair of cells $\{s,s'\}$ that lie in a common $\square_4$ block, the graph $G'[V^*\cap(s\cup s')]$ is a forest, and its components have the same vertex sets as the pair components of $\{s,s'\}$.
\end{observation}
\begin{observation}
\label{obs:connect_post_prune}
    For every edge $(u,v)\in E^*$, there is a path from $u$ to $v$ in $G'$. If $(u,v)$ is a local or a short crossing edge, this path uses only vertices in $cell(u)\cup cell(v)$.
\end{observation}

We argue that for any path in $G$, there is a path in $G^\prime$ that is ``close'' with respect to $\ell$. 

\begin{restatable}{lemma}{frechetcloseafterprune}
    \label{lem:frechetcloseafterprune}
    For any path $\pi \in G$, there exists a path $\sigma \in G^\prime$ such that $d_F(\pi, \sigma) \leq (1+4\sqrt{d})\ell$.
\end{restatable}
\begin{proof}
    We first show that for any path $\pi$ in the clustered graph $G^*$, there exists a path $\sigma$ in the pruned graph $G^\prime$ such that $d_F(\pi,\sigma) \leq 4\sqrt{d}\ell$. We then combine this with Observation~\ref{obs:frechetclosepostcluster} to conclude the lemma.
    
    Let $\pi$ be an arbitrary path in the clustered graph $G^*$. Decompose $\pi$ into a sequence of contiguous subpaths $\pi = \{\pi_1, ..., \pi_k\}$ based on the grid cell boundaries. Each subpath $\pi_i$ starts and ends at vertices $s_i$ and $t_i$ of $V^* = V^\prime$ and satisfies one of three cases. \vspace{0.2cm}
    \begin{enumerate}
        \item A maximal continuous subpath of $\pi$ that resides entirely within a single grid cell $s$ (note that such a subpath consists entirely of local edges).
        \item A short crossing edge of $\pi$. 
        \item A long crossing edge of $\pi$.
    \end{enumerate}
    
    We now show, given any path $\pi = \{\pi_1, ..., \pi_k\}$ in $G^*$, we can construct a path $\sigma = \{\sigma_1, ..., \sigma_k\}$ in the pruned graph $G^\prime$ with the same endpoints. Observe that since $G^*$ and $G^\prime$ consist of the same set of vertices and maintain connectivity (by Observation~\ref{obs:connect_post_prune}), this path always exists. We describe how this is done for each type separately.
    
    \emph{Case 1:} Since $\pi_i$ is a path in $G^*\cap s$, $\pi_i$ exists inside one of the cell components of $s$. By Observation~\ref{obs:local_forest}, the cell component was reduced to a spanning tree in $G^\prime$, which is also fully contained within $s$. Therefore, there exists a unique simple path $\sigma_i$ between $s_i$ and $t_i$ in $G^\prime$ that is fully contained within $s$. We denote this path as $\sigma_i$. Since both paths are entirely contained in a convex cell of side length $\ell$ (and $V^\prime = V^*$), it follows that $d_F(\pi_i, \sigma_i) \leq \sqrt{d}\ell$.

    \emph{Case 2:} Since $s_i$ and $t_i$ belong to cells that lie in a common $\square_4$ block the short crossing edge must be contained in a pair component. By Observation~\ref{obs:pair_forest} $s_i$ and $t_i$ are connected by a path in a spanning tree in $G^\prime$ which uses only vertices in $cell(s_i)\cup cell(t_i)$. Denote this path as $\sigma_{i}$. Since $\sigma_{i}$ must be fully contained in a $\square_4$ block, one can map all parts of $\sigma_i$ to the edge $\pi_i$ within \frechet{} distance $4\sqrt{d}\ell$.

    \emph{Case 3:} In this case the edge cannot have been removed in the process of pruning. Hence, the edge must exist in $G^\prime$. We set $\pi_i = \sigma_i$, which implies that $d_F(\pi_i, \sigma_i) = 0$ in this case.

     It follows that after constructing $\sigma$ as described above, $d_F(\pi, \sigma)\leq 4\sqrt{d}\ell$.

     Combining this with Observation~\ref{obs:frechetclosepostcluster} concludes the proof of the lemma.
\end{proof}

Before we analyse the construction time, we first prove $G^\prime$ satisfies Property~3. To do this, we show that, after pruning, there cannot be too many edges of $G^\prime$ that intersect a cube of sidelength $3\delta$.

\begin{restatable}{lemma}{nbredges}
    \label{lem:nbredges}
    Let $\ell = \frac{\varepsilon\delta}{1+4\sqrt{d}}$ be the clustering parameter of $G^\prime$. For any hypercube $C$ in $\mathbb{R}^d$ of side length $\delta$, the number of edges of $G^\prime$ that intersect $3C$ is at most $O\Big(\lambda \Big( \frac{|V\cap 27C|^{1-1/d}}{\varepsilon}+ \frac{|V\cap 27C|^{1/d}}{\varepsilon^{d-1}}\Big)\Big)$.
\end{restatable}
\begin{proof}
    From every cluster centre in $7C$, take the first $\ell/3$ units of a shortest path in $G$ to another centre. Such a path always exists because $G$ is connected. The chosen prefixes are pairwise disjoint as subsets of $G$: a common point $x$ would give $d_G(c,c') \leq d_G(c,x) + d_G(x,c') \leq 2\ell/3$ for the two centres $c,c'$, contradicting their $\ell$-separation. Since $\ell/3 < \delta$, these paths lie in $9C$. Let $k$ be the number of centres in $7C$. Then the amount of edge length of $G$ contained in $9C$ is at least $k\cdot \ell/3$. If $k=0$, the bound on $k$ below holds trivially. Otherwise, $1 \leq k \leq |V \cap 27C|$ and we apply Lemma~\ref{lem:49} to $G$ and the cube $9C$ to lower bound the number of vertices of $G$ contained in its triple expansion $27C$ to obtain:

    \begin{align*}
        |V\cap 27C| = \Omega\left(\Big(\frac{k\ell/3}{\lambda \delta}\Big)^{1+\frac{1}{d-1}}\right).
    \end{align*}
    Consequently,
    \begin{align*}
        k  = O\Big(\lambda\frac{|V\cap 27C|^{1-1/d}}{\varepsilon}\Big).
    \end{align*}

    To bound the number of edges of $G^\prime$ that overlap with $3C$ we split the edges of $E^\prime_C$ into three types, and bound each separately.
    \begin{enumerate}
        \item The edges of $E^\prime_C$ that were considered \emph{local edges} in the pruning procedure.
        \item The edges of $E^\prime_C$ that were considered \emph{short crossing edges} in the pruning procedure.
        \item The edges of $E^\prime_C$ that were considered \emph{long crossing edges} in the pruning procedure.
    \end{enumerate}
    
    We first upper bound the number of edges of types 1 and 2 in $E^\prime_C$. 
    
    A local edge has length at most $\sqrt{d}\ell$ and a short crossing edge has length at most $4\sqrt{d\ell} < \delta$. Hence, every edge of type 1 or 2 that overlaps with $3C$ has both endpoints in $5C \subseteq 7C$. By Observation~\ref{obs:local_forest}, the local edges with both endpoints in $7C$ form a forest on at most $k$ vertices. By Observation~\ref{obs:pair_forest}, for each  pair of cells in a common $\square_4$ block, the short crossing edges between them form a forest, and each cell is in a common $\square_4$ block with $O(1)$ other cells. Hence
    there can be at most $O(k)$ such edges in $E^\prime_C$.

    Next, we upper bound the number of edges of type 3. For this type we give two bounds. First, note that all edges of type 3 must be of length at least $3\ell$ in the clustered graph. For this to happen, there must exist a unique edge in the original $\lambda$-low density graph $G$ of length at least $\ell$ for each edge of type 3, as the clustering can increase edge length by at most $\ell$ for either endpoint. We will refer to these original edges in $G$ as \emph{witnesses}. Since every vertex is within graph distance $\ell$ of its cluster centre, the witness $(a,b)$ of an edge $(c_i,c_j)$ satisfies $||a-c_i|| \leq \ell$ and $||b-c_j|| \leq \ell$; in particular, $ab$ lies within distance $\ell$ of the segment $c_ic_j$, and hence intersects $4C$. Distinct edges of $G'$ have distinct witnesses.
    
    To obtain the first bound, we cover a constant expansion of $4C$ by $O((\delta/\ell)^d) = O(1/\varepsilon^d)$ cubes of side length $\ell$. We use the fact that the original graph was $\lambda$-low dense, along with Fact~\ref{fact:cube} to bound the number of such edges by $O(\lambda/\varepsilon^d)$. 
   
    The second bound splits the witnesses in $G$ into two groups. The first group consists of all witnesses that have an endpoint in $27C$. The second group consists of all witnesses that do not. For the first group, we know that since the original graph was low-dense, the maximum degree of the original graph bounds such edges by $O(\lambda \cdot|V\cap 27C|)$. Every remaining edge in the second group must be long enough for its corresponding edge in $G^\prime$ to overlap with $3C$ while both original vertices lie outside $27C$. Hence, the number of such edges is bounded by $O(\lambda)$. Our second bound is thus $O(\lambda |V\cap 27C|)$.
    
    Taking the better of these two bounds and using the fact that $\min\{x,y\} \leq x^\theta\cdot y^{1-\theta}$ with $\theta =(d-1)/d$ bounds the number of long crossing edges by
    \begin{align*}
        \min\{O(\frac{\lambda}{\varepsilon^d}),O(\lambda \cdot|V \cap 27C|)\} \leq O\big(\lambda \cdot \frac{|V \cap 27C|^{1/d}}{\varepsilon^{d-1}}\big).
    \end{align*}
    Combining the bounds on the local and crossing edges with the bound for the long crossing edges allows us to conclude the number of edges of $G^\prime$ that overlap with $3C$ is at most
    \begin{align*}
        O\Big(\lambda \Big( \frac{|V\cap 27C|^{1-1/d}}{\varepsilon}+ \frac{|V\cap 27C|^{1/d}}{\varepsilon^{d-1}}\Big)\Big).
    \end{align*}

\end{proof}

It remains to analyse the construction time and argue all the stated properties of $G^\prime$.
\begin{restatable}{lemma}{constructiontimeGprime}
    \label{lem:construction_time_G_prime}
    Given a $\lambda$-low density graph $G$ in $\mathbb{R}^d$, one can construct a simplified graph $G^\prime$ that satisfies Properties 1-3 outlined above in expected $O(\lambda|V|\log |V|)$ time.
\end{restatable}
\begin{proof}
    We have $|V^*|\leq|V|$ and $|E^*|\leq|E|\leq\lambda|V|$. Clustering the graph takes expected $O(\lambda |V| \log |V|)$ time (by Fact~\ref{fact:rnet}, plus $O(|E| + |V| \log |V|)$ time for assigning vertices to centres). We show that the pruning takes expected time $O(|V^*|+|E^*|)=O(\lambda|V|)$.
    
    \emph{Preprocessing.} We obtain the cell index of a vertex using the floor function, which takes constant time for fixed $d$. We store the indices of the non-empty cells in a hash table with at most $|V^*|$ entries, and assign each edge to the bucket of the pair of cells of its endpoints. This takes expected time $O(|V^*|+|E^*|)$.
    
    \emph{Step 1.} The cell components and their spanning trees are computed by a breadth-first search on each non-empty cell, using the local edges in the bucket of that cell. Since the cells are disjoint and every edge lies in exactly one bucket, this takes time $O(|V^*|+|E^*|)$.
    
    \emph{Step 2.} For a pair of cells $\{s,s'\}$ with bucket $X_{s,s'}$, the multigraph $H_{s,s'}$ has at most $2|X_{s,s'}|$ nodes and exactly $|X_{s,s'}|$ edges. It can be built, and a spanning forest of it computed by breadth-first search, in expected time $O(|X_{s,s'}|)$ (where the expectation comes from identifying nodes with a hash table). The buckets are disjoint, so Step 2 takes expected time $O(|E^*|)$ in total.
    
    We now prove the properties of $G^\prime$. Property~1 is satisfied by Lemma~\ref{lem:frechetcloseafterprune}, and its converse part by Observation~\ref{obs:cluster-converse}, since $G'$ is a subgraph of $G^*$. Property~2 is satisfied by the definition of $r$-nets and the fact that the pruning cannot shorten the path length between vertices. Finally, Property~3 is satisfied by Lemma~\ref{lem:nbredges}.
\end{proof}

We are now ready to bound the free space complexity between the simplified structures. 

\subsection{Bounding the approximate free space complexity}
To bound the $(1+\varepsilon)$-approximate free space complexity between the $\lambda$-low density graph $G$ and the $\lambda$-low density curve $P$, we first construct the simplified structures $G^\prime$ and $P^\prime$. We construct $P^\prime$ using Fact~\ref{fact:curveSimplification} as an $\varepsilon\delta$-simplification of $P$. We then construct $G^\prime$ as described in Section~\ref{sec:simplifyGraph} with respect to the parameter $\ell = \frac{\varepsilon\delta}{1+4\sqrt{d}}$. We will refer to $G^\prime$ as the $(\varepsilon,\delta)$-simplification of $G$. We bound the free space between the simplified structures $G^\prime$ and $P^\prime$ in order to bound the $(1+\varepsilon)$-approximate free space between the original structures. In particular, we show the free space complexity between $G^\prime$ and $P^\prime$ is at most $O\Big(\lambda^2\Big(\frac{(n|V|)^{1-1/d}}{\varepsilon^2}+\frac{n+|V|}{\varepsilon^d}\Big)\Big)$. We then apply existing techniques from~\cite{ChenDGNW11} to show one can consequently design a $(1+\varepsilon)$-approximate decider that runs in $O\Big(\lambda^2\Big(\frac{(n|V|)^{1-1/d}}{\varepsilon^2}+\frac{n+|V|}{\varepsilon^d}\Big)\log (n|E|)\Big)$ time for the map matching problem between $P$ and $G$.

\begin{restatable}{lemma}{FSComplexlowdense}
    \label{lem:FSComplex_low_dense}
    Let $G = (V,E)$ be a connected $\lambda$-low density graph, and let $P$ be a $\lambda$-low density curve of complexity $n$ in $\mathbb{R}^d$. Let $\delta>0$, and $0<\varepsilon < 1$. Let $G^\prime$ be an $(\varepsilon,\delta)$-simplification of $G$, and let $P^\prime$ be an $(\varepsilon\delta)$-simplification of $P$. The free space complexity with respect to $\delta$ between $G^\prime$ and $P^\prime$ (denoted as $FS_{\delta}(P^\prime, G^\prime)$) is at most $O\Big(\lambda^2\Big(\frac{(n|V|)^{1-1/d}}{\varepsilon^2}+\frac{n+|V|}{\varepsilon^d}\Big)\Big)$.
\end{restatable}
\begin{proof}
    We bound the number of vertex-edge pairs in the free-space diagram of $(P^\prime, G^\prime)$ that have free space, as follows. The running time of the decider of~\cite{ChenDGNW11} depends on the number of cells of the free space surface, each corresponding to an edge of $P’$ and an edge of $G’$, that intersect the relevant free space. Apart from the cells containing a starting point, a cell can only be reached through a non-empty free boundary, which corresponds to a vertex of one structure and an edge of the other at distance at most $\delta$. The boundary of a vertex of $P’$ is shared by at most two cells, and the boundary of a vertex $v$ of $G’$ by $\deg_{G’}(v)$ cells. We therefore count each pair consisting of a vertex of $P’$ and an edge of $G’$ twice, and each pair consisting of a vertex $v$ of $G’$ and an edge of $P’$ exactly $\deg_{G’}(v)$ times; the resulting weighted count bounds $FS_{\delta} (P’,G’)$. 
    Let
    \begin{align*}
        B = \lambda^2 \Big(\frac{(n|V|)^{1-1/d}}{\varepsilon^2}+\frac{n+|V|}{\varepsilon^d}\Big).
    \end{align*}
    We partition the pairs into two types, according to whether the vertex in each pair belongs to $P^\prime$ or $G^\prime$. We charge each pair to the cell containing this vertex. We then bound the number of pairs of each type separately, showing that each type contributes at most $O(B)$ pairs.

    Consider a $\delta$-grid overlaying the two simplified structures. For each grid cell $C$, let $V_P(C)$ and $V_G(C)$ denote the sets of vertices of $P$ and $G$, respectively, that overlap $C$. Similarly, let $E_P(C)$ and $E_G(C)$ denote the sets of edges of $P$ and $G$, respectively, that overlap $C$.  Recall that $xC$ denotes the cube obtained by scaling $C$ by a factor of $x$ about its centre.

    We first consider the number of charges to vertices of $P^\prime$ in $C$ by edges of $G^\prime$ in $3C$. Note that all edges of $G^\prime$ that are within distance $\delta$ of a vertex of $V_{P^\prime}(C)$ must overlap with $3C$. From Lemma~\ref{lem:nbredges} we obtain: 
    \begin{align}
        \label{eq:edgesGRd}
        |E_{G^\prime}(3C)| = O\Big(\lambda\Big(\frac{|V_G(27C)|^{1-1/d}}{\varepsilon}+\frac{|V_G(27C)|^{1/d}}{\varepsilon^{d-1}}\Big)\Big).
    \end{align}
    Next, we bound the number of vertices of $P^\prime$ that lie in $C$.
    Recall that every edge of $P^\prime$, except possibly the last, has length at least $\varepsilon\delta$. For each vertex of $V_{P^\prime}$ other than the final vertex, we charge the first $\varepsilon\delta/2$ units of the subsequent subcurve of $P$ to that vertex. These subcurves are parameter-disjoint and lie entirely within $5C$. We may assume $V_{P'}(C) \neq \emptyset$; since the vertices of $P'$ are vertices of $P$, then $|V_P(15C)| \geq 1$. Since $P$ is $\lambda$-low dense, applying Lemma~\ref{lem:49} to $5C$ yields
    \begin{align*}
       |V_P(15C)| = \Omega\left(\Big(\frac{ |V_{P^\prime}(C)|\varepsilon\delta}{\lambda\delta}\Big)^{1+\frac{1}{d-1}}\right).
    \end{align*}
    Consequently,
    \begin{align}
    \label{eq:verticesPprime}
        |V_{P^\prime}(C)| = O\Big(\lambda \frac{|V_P(15C)|^{1-1/d} }{\varepsilon}\Big).
    \end{align}
    Using Equations~(\ref{eq:edgesGRd}) and~(\ref{eq:verticesPprime}) we can bound the charges incurred in $C$ by pairs that consist of a vertex of $P^\prime$ and an edge of $G^\prime$ as follows:
    \begin{align*}
        |V_{P^\prime}(C)|\cdot|E_{G^\prime}(3C)| = 
    \end{align*}
    \begin{align}
    \label{eq:pairs1Rd}
        O\left(\lambda^2\left(\frac{(|V_P(15C)| \cdot |V_G(27C)|)^{1-1/d}}{\varepsilon^2}+\frac{|V_P(15C)|^{1-1/d} \cdot |V_G(27C)|^{1/d}}{\varepsilon^d}\right)\right).
    \end{align}
    Observe that the expanded cells have bounded overlap. Therefore, $\sum_C|V_P(15C)| =  O(n)$,
   and
       $\sum_C|V_G(27C)| =  O(|V|)$.
    Using the fact that $2(1-1/d) \geq 1$, Cauchy-Schwarz and Hölder's inequality give
    \begin{align*}
        \sum_C (|V_P(15C)| \cdot |V_G(27C)|)^{1-1/d} = O((n|V|)^{1-1/d}), 
    \end{align*}
    and
    \begin{align*}
        \sum_C |V_P(15C)|^{1-1/d} \cdot |V_G(27C)|^{1/d} = O(n^{1-1/d}|V|^{1/d}) = O(n+|V|).
    \end{align*}
    Substituting these bounds into Equation~(\ref{eq:pairs1Rd}), we conclude that the total number of pairs of the first type is $O(B)$.

    Next, we bound the number of pairs of the second type, in which the vertex belongs to $V_{G^\prime}(C)$ and the edge belongs to $E_{P^\prime}(3C)$. By Equation~(\ref{eq:edgesGRd}), we obtain that
    \begin{align}
    \label{eq:VerticesGPrime}
        \sum_{v\in V_{G'}(C)} \deg_{G'}(v)  
       \leq 2\cdot|E_{G^\prime}(3C)| = O\Big(\lambda\Big(\frac{|V_G(27C)|^{1-1/d}}{\varepsilon}+\frac{|V_G(27C)|^{1/d}}{\varepsilon^{d-1}}\Big)\Big).
    \end{align}
    It remains to bound the number of edges of $P^\prime$ that overlap $3C$. Recall that every edge of $P^\prime$, except possibly the last, has length at least $\varepsilon\delta$. We may assume $\varepsilon \leq  1/4$. Let $e' = p_ip_j$ be an edge of $E_{P'}(3C)$ other than the last, let $x\in e'\cap 3C$, and let $\sigma \subseteq e'$ be a subsegment containing $x$ of length $\min\{|e'|, \delta\}$; then $\sigma \subseteq 5C$. Since $d_F(P[p_i,p_j], e') \le \varepsilon \delta \le \delta/4$, the segment $\sigma$ is matched to a subcurve of $P[p_i,p_j]$ that lies in $6C$. If $\sigma = e'$, this subcurve is $P[p_i, p_j]$ itself, of length at least $|e'| \ge \varepsilon \delta$; otherwise its length is at least $\delta - 2\varepsilon\delta \ge \delta/2 \ge \varepsilon \delta/2$. Hence, the subcurve of $P$ corresponding to $e'$ contributes at least $\varepsilon \delta/2$ to $6C$.

    We charge the first $\varepsilon\delta/4$ units of this contribution to each corresponding subcurve. Since the distinct edges of $P^\prime$ represent parameter-disjoint subcurves of $P$, the charged portions are parameter-disjoint. It follows that the total length of $P$ inside $6C$ is at least $(|E_P^\prime(3C)| -1)\cdot \varepsilon\delta/4$. Applying Lemma~\ref{lem:49} to $6C$ yields, if $V_P(18C) \neq \emptyset$,
    \begin{align*}
    |V_P(18C)| = 
        \Omega \left(\Big( \frac{|E_{P^\prime}(3C)|\varepsilon\delta}{\lambda \delta} \Big)^{1+\frac{1}{d-1}} \right).
    \end{align*}
    Consequently,
    \begin{align}
        \label{eq:edgesPprime}
        |E_{P^\prime}(3C)| = O\Big(\lambda\frac{|V_P(18C)|^{1-1/d}}{\varepsilon}\Big).
    \end{align}
    If $V_P(18C) = \emptyset$, the first statement of Lemma~\ref{lem:49} gives $|E_{P'}(3C)| = O(\lambda/\varepsilon)$ instead. Only cubes with $V_{G'}(C) \neq \emptyset$ contribute charges of the second type, and there are at most $|V|$ of them, so by Equation~(\ref{eq:VerticesGPrime}) these cubes contribute 
    $O(\frac{\lambda^2}{\varepsilon} \sum_C(( \frac{|V_G(27C)|^{1-1/d}}{\varepsilon}+\frac{|V_G(27C)|^{1/d}}{\varepsilon^{d-1}}))= O(B)$
    in total. For the remaining cubes Equation~(\ref{eq:edgesPprime}) holds.

    Combining Equations~(\ref{eq:VerticesGPrime}) and~(\ref{eq:edgesPprime}), we obtain the following bound on the total number of charges.
    \begin{align*}
    \label{eq:pairs2Rd}
        |E_{P^\prime}(3C)| \cdot \sum_{v\in V_{G'}(C)} \deg_{G'}(v)
        = O\Big(\lambda^2 \frac{|V_P(18C)|^{1-1/d}}{\varepsilon} \cdot \Big( \frac{|V_G(27C)|^{1-1/d}}{\varepsilon}+\frac{|V_G(27C)|^{1/d}}{\varepsilon^{d-1}}\Big)\Big).
    \end{align*}
    Expanding the equation, using Cauchy-Schwarz and Hölder's inequality, and using $x^{1-1/d} \leq x$ for all $x \geq 1$ yields
    \begin{align*}
        \sum_{C} |E_{P^\prime}(3C)| \cdot \sum_{v\in V_{G'}(C)} \deg_{G'}(v)= O(B).
    \end{align*}
    We thus conclude the second type of pairs contributes $O(B)$ pairs in total.
    
   Note that a vertex of each original structure can be charged only a constant number of times because it belongs to only a constant number of fixed expansions of grid cells. Therefore, the free-space complexity between the two simplified structures is at most
    \begin{align*}
        O(B) = O\Big(\lambda^2 \Big(\frac{(n|V|)^{1-1/d}}{\varepsilon^2}+\frac{n+|V|}{\varepsilon^d}\Big)\Big),
    \end{align*}
    as claimed.
\end{proof}

For $d=2$ dimensions, using the inequality $\sqrt{n|V|} \leq \frac{n+|V|}{2}$, the approximate free space complexity bound simplifies to $O(\frac{\lambda^2}{\varepsilon^2}(n+|V|))$.

\subsection{A $(1+\varepsilon)$-approximate Decider}\label{sec:approxDeciderLowDense}
Let $FS_\delta(P,G)$ denote the cells with relevant free space in the free space surface w.r.t parameter $\delta$ between $P$ and $G$. The authors of~\cite{ChenDGNW11} show that for a polygonal curve $P$ of complexity $n$ and graph $G$ with $|V|$ vertices and $|E|$ edges in $\mathbb{R}^d$, there exists an algorithm
\textsc{Decider}$(P,G,\delta)$ that decides whether there is a path $\pi\in G$ such that $d_F(\pi,P)\le\delta$ in
$O(|V|+FS_\delta(P,G) \log(n|E|))$ time and $O(|V|+FS_\delta(P,G))$ space. If such a $\pi$ exists, the
algorithm also returns it. We apply this decider to the simplified structures to obtain the following lemma.

\begin{restatable}{lemma}{freespaceToDecider} \label{lem:freespaceToDecider} Let $P$ be a polygonal curve of complexity $n$ and let $G=(V,E)$ be an arbitrary graph of complexity $|V|+|E|$. Let $\delta>0$ and $0<\varepsilon<1$. Let $G'$ be an $(\varepsilon,\delta)$-simplification of $G$, and let $P'$ be an $(\varepsilon\delta)$-simplification of $P$. There exists an algorithm \textsc{ApproxDecider} that decides in $O(|E|+FS_{(1+2\varepsilon)\delta}(P',G')\log(|E|n))$ time whether
there exists a path $\pi\in G$ such that $d_F(\pi,P)\le(1+4\varepsilon)\delta$ or whether, for all paths $\pi\in G$,
$d_F(\pi,P)>\delta$.
\end{restatable}
\begin{proof}
Let $\delta_1=(1+2\varepsilon)\delta$. \textsc{ApproxDecider} runs \textsc{Decider}$(P',G',\delta_1)$ and returns its answer, which takes
$O(|E|+FS_{\delta_1}(P',G')\log(|E|n))$ time. If there is a path $\pi\in G$ with $d_F(\pi,P)\le\delta$, then by
Property~1 there is a path $\sigma\in G'$ with $d_F(\pi,\sigma)\le(1+4\sqrt d)\ell=\varepsilon\delta$, and $d_F(P,P')\le\varepsilon\delta$ by Fact~\ref{fact:curveSimplification}, so
$d_F(\sigma,P')\le\delta_1$ and the decider answers YES. Conversely, if the decider returns a path $\sigma\in G'$ with $d_F(\sigma,P')\le\delta_1$, then by Observation~\ref{obs:cluster-converse} (as $G'$ is a subgraph of $G^*$) there is a path $\pi\in G$ with $d_F(\pi,\sigma)\le\ell\le\varepsilon\delta$, and
$d_F(\pi,P)\le d_F(\pi,\sigma)+d_F(\sigma,P')+d_F(P',P)\le(1+4\varepsilon)\delta$.
\end{proof}

We combine the results from Lemmas~\ref{lem:construction_time_G_prime},~\ref{lem:FSComplex_low_dense}, and~\ref{lem:freespaceToDecider} to conclude Theorem~\ref{thm:low_density}. We apply Lemma~\ref{lem:freespaceToDecider} with $\varepsilon/4$, and write $\varepsilon$ for $\varepsilon/4$ in what follows.
It remains to bound $FS_{\delta_1}(P',G')$ for $\delta_1=(1+2\varepsilon)\delta$. Let $\varepsilon_1=\varepsilon\delta/\delta_1\in[\varepsilon/3,\varepsilon]$. The simplifications depend only on the products $\ell=\frac{\varepsilon\delta}{1+4\sqrt d}=\frac{\varepsilon_1\delta_1}{1+4\sqrt d}$ and $\varepsilon\delta=\varepsilon_1\delta_1$, so $G'$ is also an $(\varepsilon_1,\delta_1)$-simplification of $G$ and $P'$ is an $(\varepsilon_1 \delta_1)$-simplification of $P$. Lemma~\ref{lem:FSComplex_low_dense}, applied with $\delta_1$ and $\varepsilon_1$, gives $FS_{\delta_1}(P',G')=O(B)$. Constructing $P'$ takes $O(n)$ time and constructing $G'$ takes expected $O(\lambda|V|\log|V|)$ time; as $\lambda\ge 1$, $|E|\le\lambda|V|$ and $\log |V|=O(\log (n|E|))$, both terms, as well as the $O(|E|)$ term of Lemma~\ref{lem:freespaceToDecider}, are dominated by $O(B\log(n|E|))$.

\section{From a Decider to an Approximation Algorithm}\label{app:decider-to-approximation}

In this section, we turn the deciders of Theorems~\ref{thm:decisionC}
and~\ref{thm:low_density} into approximation algorithms for
Problem~\ref{problem:1}. Recall that $P$ is a polygonal curve with $n$
vertices, $G=(V,E)$ is a connected graph, and $r^*=\min_\pi d_F(P,\pi)$ is
the optimal value of Problem~\ref{problem:1}. Since the deciders only
approximately compare $r^*$ with a given radius, we need a list of
candidate radii to search.

\begin{theorem}\label{thm:approximation}
Given a curve $P$ and a graph $G=(V,E)$, and fixed constants $\varepsilon$
and $d$, one can compute a $(1+\varepsilon)$-approximation to
$r^*=\min_\pi d_F(P,\pi)$, with running time
$
  \widetilde O(cn+c^2|E|)
$
if the curve $P$ is $c$-packed, and with expected running time
$
  \widetilde O\big(\lambda^2\big((n|V|)^{1-1/d}+n+|V|\big)\big)
$
if the curve $P$ is $\lambda$-low-dense and the graph $G$ is
$\lambda$-low-dense.
\end{theorem}

\begin{proof}
We follow the approach of Driemel, Har-Peled and
Wenk~\cite[Section~3]{DriemelHW12}. In particular, we first turn the
fuzzy decider into a complete one, then use it to search over vertex
distances to fix a simplification of $P$ and $G$, and finally search for critical values over the simplifications.

\textbf{Step 1: A complete decider.}
The deciders of Theorems~\ref{thm:decisionC} and~\ref{thm:low_density}
are fuzzy: at radius $s$ with accuracy $\varepsilon/8$, a \textsc{yes}
certifies that $r^*\le(1+\varepsilon/8)s$, and a \textsc{no} certifies
that $r^*>s$. We want to perform a binary search over candidate radii to find $r^*$, but when $r^*$ is close to $s$, the two answers do not tell us on which side of $s$ the optimum lies. We therefore build
a complete decider that, given $t>0$, returns either (i) $r^*\le t$,
(ii) $r^*>t$, or (iii) a $(1+\varepsilon)$-approximation to $r^*$.

To do so, we run the fuzzy decider at $t/(1+\varepsilon/8)$ and at $t$
\cite[Section~6]{GudmundssonSW24}. If the first call returns \textsc{yes},
we return (i). If the second call returns \textsc{no}, we return (ii).
Otherwise $t/(1+\varepsilon/8)<r^*\le(1+\varepsilon/8)t$, so
$(1+\varepsilon/8)t\le(1+\varepsilon/8)^2r^*\le(1+\varepsilon)r^*$, and we
return $(1+\varepsilon/8)t$ as (iii). Whenever the complete decider returns (iii), we stop.

\textbf{Step 2: Finding the simplification.}
We use the complete decider in a binary search over candidate radii
derived from the vertex distances. The goal is to either obtain (iii), or
find two consecutive candidate radii $L\le r^*\le U$ such that no vertex
distance lies between a small fraction of $L$ and a large multiple of $U$.
Let $N=n+|V|$. From a Euclidean WSPD we obtain $O(N)$ values $z$ such that
every positive distance between two input vertices lies in $[z/2,2z]$ for
some $z$~\cite[Lemma~3.8]{ChenDGNW11}. For each $z$, we sample
$[z/16,32z/\varepsilon]$ geometrically, including both endpoints, with
consecutive ratios at most $1+\varepsilon/8$. We binary search all these values
together, using zero and the bounding-box diagonal as initial bounds.

After $O(\log N)$ calls to the complete decider, we either have stopped by returning (iii) or have
consecutive values $L\le r^*\le U$. If $U\le(1+\varepsilon/8)L$, we
return $U$, since $U\le(1+\varepsilon/8)r^*$. Otherwise $U>(1+\varepsilon/8)L$, so no
interval $[z/16,32z/\varepsilon]$ overlaps with $(L,U)$: its samples are at most a
factor $1+\varepsilon/8$ apart, and none of them lies strictly between the
consecutive values $L$ and $U$. Hence $32z/\varepsilon\le L$ or
$z/16\ge U$ for every $z$. Since every vertex
distance lies in some $[z/2,2z]$, it is either at most
$2z\le\varepsilon L/16$ or at least $z/2\ge 8U$. We use this property to obtain the required simplification.

We now simplify $P$ and $G$ by clustering their vertices. We cluster
the input vertices, of both $P$ and $G$, so that two vertices are in the
same cluster if and only if they are at distance at most
$\varepsilon L/16$. This clustering exists: if $u,v$ and $v,w$ are at
distance at most $\varepsilon L/16$, then $u,w$ are at distance at most
$\varepsilon L/8<8U$, and hence, by the property above, at most
$\varepsilon L/16$. Using a grid, we compute the clusters in
$\widetilde O(N)$ time. We fix a
cluster centre in each cluster and move every vertex to the centre of its
cluster. The simplified curve $\bar P$ is the resulting curve, with
consecutive repeated vertices removed. The simplified graph $\bar G$ is
the resulting graph, with every edge inside a cluster contracted, so each
vertex of $\bar G$ replaces a connected set of vertices of $G$ in one
cluster.

The simplification changes the Fr\'echet distance by at most
$\varepsilon L/16$, since every vertex moves by at most $\varepsilon L/16$,
and every path in $\bar G$ lifts to a path in $G$ via the clustering.
For the remainder of this proof, we work with $\bar P$ and $\bar G$. Let
$\bar r=\min_{\bar\pi}d_F(\bar P,\bar\pi)$ be the Fr\'echet distance
between $\bar P$ and the closest path $\bar\pi$ in $\bar G$. Note that
$|\bar r-r^*|\le\varepsilon L/8$.

\textbf{Step 3: Searching the critical values.}
It remains to approximate $\bar r$. For this, we find a list of critical
values that contains $\bar r$. By the critical-value argument of Alt and
Godau~\cite{AltG95} (see~\cite[Section~2.6]{AltERW03} for map matching,
and~\cite[Section~2.2.1]{DriemelHW12} for a description of the events),
$\bar r$ is one of the following:
(a) the distance between two endpoints,
(b) the distance between a vertex and a segment, or
(c) the distance from two vertices to a common point on a segment.
We show that cases (a) and (c) reduce to case (b). Since simplified
vertices are at least $8U$ apart and $\bar r\le r^*+\varepsilon L/8<2U$,
any two simplified vertices at distance at most $2\bar r$ coincide. In
case (a), the two vertices are at distance $\bar r$, so they coincide and
$\bar r=0$. In case (c), the two vertices are at distance at most
$2\bar r$, so they coincide, and $\bar r$ is the distance between this
vertex and the segment. Therefore, we only need to consider case (b),
that is, $\bar r$ is the distance between a simplified vertex and a
segment.

How we search for $\bar r$ depends on the input model. For a low-density
curve and graph, we binary search the vertex--segment distances in the
reachable free space. For a $c$-packed curve, we decide whether a directed
graph contains a directed path, as in~\cite[Section~2.2]{GudmundssonMW24},
and binary search the radius.

\emph{A low-density curve and graph.}
We enumerate the reachable free-space cells at radius $2U$ and collect the
vertex--segment distances in case (b) that occur on the sides of these
cells. We only keep the vertex--segment distances that are at most $2U$;
these include $\bar r$, as shown above. We binary search these distances with the
exact decider of Chen et al.~\cite{ChenDGNW11}, which works for any curve
and graph (see Section~\ref{sec:approxDeciderLowDense}), to obtain $w=\bar r$. Since $|\bar r-r^*|\le\varepsilon L/8$
and $L\le r^*$, we have
$r^*\le w+\varepsilon L/8\le r^*+\varepsilon L/4\le(1+\varepsilon)r^*$, so
$w+\varepsilon L/8$ is a $(1+\varepsilon)$-approximation to $r^*$, as
required.

Finally, we analyse the running time. Each call to the complete decider
of Step~1 makes two calls to the decider of
Theorem~\ref{thm:low_density}. In Step~2, we make $O(\log N)$ such calls,
which take $\widetilde O\big(\lambda^2\big((n|V|)^{1-1/d}+n+|V|\big)\big)$
expected time in total, and computing the candidate radii and the
clustering takes $\widetilde O(N)$ time. In Step~3, the enumeration and
each call to the exact decider take time near-linear in $n+|V|$ plus the
number of reachable cells. To bound the number of cells, we charge each
cell to the shorter of its two edges, of length $\ell$. The edge of $P$ or
$G$ that the other edge comes from has length at least $\ell/2$ and
intersects a ball of radius $2\ell$ about the midpoint of the shorter
edge. Low density allows only $O_d(\lambda)$ such edges, so each edge is
charged $O_d(\lambda)$ times, and there are
$O_d(\lambda(n+|E|))=O_d(\lambda^2(n+|V|))$ cells in total, since
$|E|\le\lambda|V|$. Hence the expected running time is
$\widetilde O\big(\lambda^2\big((n|V|)^{1-1/d}+n+|V|\big)\big)$.

\emph{A $c$-packed curve.}
For each vertex $v$ of $\bar G$, we use
Lemma~\ref{lem:LayeredGraphVertices} to find the segments of $\bar P$
intersecting $B(v,2U)$. There are $O(c)$ such segments, since their lengths
are at least $8U$ and $\bar P$ is still $O(c)$-packed~\cite[Lemma~4.3]{DriemelHW12}. Next, we build a directed graph with a node $(v,x)$ for each projection of a
vertex $v$ of $\bar G$ onto a segment reported by
Lemma~\ref{lem:LayeredGraphVertices}, where $x$ is its
parameter on $\bar P$. For each oriented edge $uv$ of $\bar G$ and nodes
$(u,x),(v,y)$ with $x\le y$, we use
Fact~\ref{fact:jaywalkingDataStructure}~\cite{jaywalking} to compute a
$(1+\varepsilon/8)$-approximation $D'$ of $D=d_F(\bar P[x,y],uv)$. For a
radius $t$, we insert the directed edge from $(u,x)$ to $(v,y)$ if
$D'\le t$, and use breadth-first search to decide whether there is a
directed path from a node $(v,x)$ whose vertex $v$ coincides with the
first vertex of $\bar P$ and whose parameter $x$ is the start of
$\bar P$, to a node whose vertex coincides with the last
vertex of $\bar P$ and whose parameter is the end of $\bar P$. We binary search the values $D'$ for the smallest $t=w$ for
which there is such a path. Concatenating the matchings along this path
gives a matching between $\bar P$ and a path in $\bar G$ in which every
matched pair is at distance at most $w$, so $w\ge\bar r$. Moreover,
$w\le(1+\varepsilon/8)\bar r$, since in an optimal matching we can move
the point of $\bar P$ paired with each vertex of $\bar\pi$ to the
projection of that vertex, as simplified vertices are far apart compared
with $\bar r$. The resulting matching pairs each vertex $v$ of $\bar\pi$
with a point $x$ such that $(v,x)$ is a node, and between consecutive
such nodes it has $D\le\bar r$, so these nodes form a directed path for
$t=(1+\varepsilon/8)\bar r$.

Therefore, $w$ is a $(1+\varepsilon/8)$-approximation to $\bar r$.
Since $|\bar r-r^*|\le\varepsilon L/8$ and $L\le r^*$, we have
$r^*\le w+\varepsilon L/8\le(1+\varepsilon/8)(r^*+\varepsilon L/8)+\varepsilon L/8\le(1+\varepsilon)r^*$,
so $w+\varepsilon L/8$ is a $(1+\varepsilon)$-approximation to $r^*$, as
required.

Finally, we analyse the running time. Each call to the complete decider
of Step~1 makes two calls to the decider of
Theorem~\ref{thm:decisionC}. In Step~2, we make $O(\log N)$ such calls,
which take $\widetilde O(cn+c^2|E|)$ time in total, and computing the
candidate radii and the clustering takes $\widetilde O(N)$ time. In Step~3, there are
$O(c^2|E|)$ directed edges, each value $D'$ takes polylogarithmic time to
compute, and the binary search makes a logarithmic number of calls to
breadth-first search. Hence the running time is
$\widetilde O(cn+c^2|E|)$.
\end{proof}

\section{Concluding Remarks}
In this paper, we presented two algorithms for the map matching problem that relax the structural assumptions on the input graph and curve made by~\cite{ChenDGNW11}. First, we showed that assuming $c$-packedness of the curve alone suffices to obtain a subquadratic time approximation algorithm, with no assumption on the graph. This extends the results of~\cite{GudmundssonMW24} and~\cite{DBLP:conf/sosa/ConradiHHO26} to map matching. Second, we obtained a subquadratic time approximation algorithm for the case in which both the graph and the curve satisfy $\lambda$-low density. This extends the result of~\cite{DriemelHW12} for two $\lambda$-low density curves to map matching. Both results improve on~\cite{ChenDGNW11}: the first removes the assumption on the graph~$G$, and the second generalises the assumption on the curve~$P$.

We conclude with two open problems. First, can we improve the dependency on $c$ to linear for map matching a $c$-packed curve to a general graph? The current $c^2$ factor stems from searching the directed graph $H$ for a \emph{valid} path. One question is thus whether this part can be improved or whether a new approach can improve this. One approach that may seem natural is  to extend the results by the authors of~\cite{DBLP:conf/sosa/ConradiHHO26}, who design subquadratic time approximation algorithms for computing the \frechet{} distance between two curves when only one curve is $c$-packed. The authors take a different approach to that in~\cite{GudmundssonMW24}, and show that if the \frechet{} distance is close to a given guess, the free space complexity must be small. To argue this, they rely on the fact that the two structures must be entirely and monotonically mapped to one another. They use this fact to argue that long edges of the general structure must satisfy some approximate c-packedness in order to be sufficiently close to the $c$-packed curve. In map matching however, the entire graph does not need to be mapped to the curve. Instead, only a single path in the graph is required to map to the curve.  Hence, unless the path is already known, this approach does not directly extend to map matching. As such, bounding the free space complexity between a general graph and $c$-packed curve may not be possible.

Second, is $\lambda$-low density a sufficiently strong assumption for one-sided map matching? In the literature, $c$-packed graphs have often been used as a first step to obtaining good bounds for $\lambda$-low density graphs. A natural question to ask is thus whether, similar to the case of $c$-packedness, we can remove the $\lambda$-low density assumption on the graph, or whether $\lambda$-low density is not a sufficiently strong assumption for one-sided results in map matching. Results in one-sided map-matching for $c$-packedness are heavily inspired by results for one-sided $c$-packedness when computing the \frechet{} distance between curves. A natural first step would thus be to consider whether it is possible to obtain subquadratic time algorithms for approximating the \frechet{} distance between two curves when only one curve satisfies $\lambda$-low density.

\section{AI statement}

All ideas in this paper are solely due to the authors, obtained without the use of LLMs. Only in the final stages was Claude 5.5 used to assist with writing, polishing and identifying errors.

\bibliographystyle{plainurl}
\bibliography{main}
\end{document}